\documentclass[12pt,preprint]{emulateapj}

\shorttitle{NLTT White Dwarfs}
\shortauthors{Kawka \& Vennes}

\begin{document}

\title{Spectroscopic Identification of Cool White Dwarfs in the Solar Neighbourhood}

\author{Adela Kawka}
\affil{Astronomick\'y \'ustav AV {\v C}R, Fri{\v c}ova 298, 251 65 Ond{\v r}ejov, Czech Republic}

\and 

\author{St\'ephane Vennes}
\affil{Department of Physics and Space Sciences, Florida Institute of Technology, 150 W University Blvd., Melbourne, FL 32901-6975}

\begin{abstract}
The New Luyten Two-Tenths catalog contains a large number of high-proper
motion white dwarf candidates that remain to be spectroscopically confirmed.
We present new spectroscopic observations as well as SDSS archival spectra of 49 white dwarf candidates 
which have been selected from the revised NLTT catalog of \citet{sal2003}.
Out of these, 34 are cool DA white dwarfs with temperatures ranging
from approximately 5000 K up to 11690 K, and 11 are DC white dwarfs
with temperatures ranging from 4300 K (NLTT~18555) up to 11000 K.
Three of the DA white dwarfs also display abundances of heavy elements
(NLTT 3915, NLTT 44986 and NLTT 43806) and one is a cool magnetic white dwarf (NLTT 44447)
with an estimated magnetic field strength of 1.3 MG.
We also present a new cool DQ white dwarf (NLTT 31347) with an estimated
temperature of 6250 K. We supplement our sample with 
SDSS {\it ugriz} photometry for a fraction of 
the newly identified white dwarfs. A kinematical study of this sample of white 
dwarfs, characterized by proper motions ranging from 0.136 to 0.611$\arcsec$yr$^{-1}$ suggest that they belong to the thin disk population.
\end{abstract}


\section{Introduction}

The current census of white dwarfs in the solar neighborhood is believed to 
be complete only to about 13 pc, and is measurably incomplete within
20 pc \citep{hol2002,sch2004}. The local sample of white dwarfs appears to be
an old population consisting of mostly cool white dwarfs. 
The average absolute magnitude for the local census of white dwarfs is
$M_V = 13.7$ which corresponds to a temperature of $\sim 7000$ K.
A significant
fraction of white dwarfs within 13 pc \citep[$20\pm8\%$]{kaw2003} show the presence of a magnetic field.
\citet{hol2002}
determined that over a quarter of the white dwarfs within 20 pc are in binary 
systems. Based on recent literature,
we updated the spectral classifications of the sample of white dwarfs within 20 pc
\citep{hol2002} and found that 
about 70\% of the stars are hydrogen-rich.
The remaining 30\% are helium-rich, and out of these helium rich white dwarfs
about half are DC white dwarfs, about a third are cool DQ white dwarfs, while
the remainder are DZ white dwarfs. 

The New Luyten Two-Tenths (NLTT) catalog has been used to search 
for nearby cool dwarfs and subdwarfs \citep{giz1997,rei2003,yon2003,rei2005} 
and for white dwarfs \citep{lie1977,hin1986,ven2003,kaw2004,kaw2005}. 
However, a large number of objects still require a formal spectroscopic 
confirmation. Recently, \citet{sal2003} have revised the coordinates and proper motions of
most stars in the NLTT catalog by cross-correlating the catalog with the 
Two Micron All Sky Survey (2MASS) and the USNO-A catalogs. One problem with 
the original catalog was that there was no obvious separation between 
different populations because the R and B photographic bands do not provide
a broad enough baseline to distinguish between main-sequence, subdwarf and
white dwarf populations. \citet{sal2002} showed that using an optical/infrared ($V-J$)
reduced proper motion diagram they could distinguish between these groups of
stars and from this diagram they listed 23
white dwarf candidates within 20 pc from the Sun.
\citet{ven2003} and \citet{kaw2004} found
that only a third of these nearby candidates are white dwarfs, with the remainder being cool
red dwarfs and subdwarfs.

Recently, several projects have been initiated to search for high-proper
motion stars, such as the SuperCOSMOS-RECONS survey \citep{ham2004} and LSPM 
Catalog \citep{lep2005} which utilizes the SuperCOSMOS Sky Survey 
\citep{ham2001} and SUPERBLINK \citep{lep2002}, respectively. These surveys
have detected many high-proper motion stars which are either nearby or
have been labeled as halo candidates. One such object is
the cool white dwarf WD 0346$+$246 \citep{ham1997}, which has definite halo
kinematics and is very cool ($T\sim3800$ K) and hence old enough to belong to 
the halo.
A total of 299
new stars with $0.4\arcsec$ yr$^{-1}$ have been 
discovered using the SuperCosmos-RECONS survey \citep{sub2005}, out of which 148 
have proper motions greater than $0.5\arcsec$yr$^{-1}$ and 43 are within 25 pc from
the Sun. \citet{lep2002,lep2003} reported the discovery of 198 new stars with
proper motions greater than $0.5\arcsec$yr$^{-1}$ using SUPERBLINK. Their survey also
included the nearby halo white dwarf PM J13420$-$3415 \citep{lep2005b}.

The NLTT catalog appears to be complete
at Galactic latitudes $|b| > 15^\circ$, however at low Galactic latitudes
the catalog appears to be significantly incomplete. On the other hand, \citet{sal2003}
notes that the completeness for white dwarfs is very high, i.e., the 
coverage of white dwarfs appears to be completely uniform. The reason for this
may be that Luyten concentrated on blue objects, which are no more common in 
the plane than elsewhere, and hence Luyten was able to detect most 
high-proper motion white dwarfs above $\delta > -32.5\degr$.
Another survey to search for cool
white dwarfs with high-proper motion using the Guide Star Catalog II was 
initiated by \citet{car2005} with the aim of placing constraints on the 
halo white dwarf space density. Their search resulted in the discovery of 24
new white dwarfs.
Many earlier proper motion surveys, in 
particular the
Luyten proper motion surveys relied on plate pairs taken a
decade or more apart, and very high-proper motion stars were most likely lost 
in the background.
A search for ultra-high proper motion stars was 
performed by \citet{tee2003} using the SkyMorph database of the Near Earth
Asteroid Tracking (NEAT) project \citep{pra1999} and found a main-sequence star
with spectral type M6.5 and with a proper motion of 
$5.05\arcsec\pm0.03\arcsec {\rm yr}^{-1}$ at a distance of 
$2.4^{+0.7}_{-0.4}$ pc which they obtained from the trigonometric parallax.

Since most of the stars in the NLTT catalog are brighter than 19th magnitude, 
we do not expect to find many halo white dwarf candidates. Since the halo
star formation history is assumed to be a burst occurring 12 Gyrs ago, then
most of the halo white dwarfs would be required to have cooling ages greater
than 10 Gyrs (and hence $M_V \ga 16$). Given that the percentage of halo white 
dwarfs in the local population is $\sim 2\%$ \citep{pau2005}, then most halo candidates
are likely to be at large distances and hence too faint to be included in the
NLTT catalog. 

In this paper, we pursue our study of white dwarf candidates 
\citep{kaw2004} from the revised NLTT (rNLTT) catalog of \citet{sal2003}.
Some preliminary results were presented in \citet{kaw2005}.
We present the list of white dwarfs observed in \S 2 and their 
spectroscopic observations in \S 2.1 with complementary optical and infrared
photometric data presented in \S 2.2. The model atmospheres and spectra we 
used to analyze our data are described in \S 3 and in \S 4 we present our
analysis of the new white dwarf stars. We discuss our findings in \S 5 and
summarize in \S 6.

\section{Observations of White Dwarf Candidates}

Table~\ref{tbl1} lists the white dwarf candidates for which we have obtained
a spectroscopic identification. 
Many of the white dwarfs were listed in the Luyten White Dwarf Catalogs (LWDC)
\citep{luy1970,luy1977}, and required spectroscopic confirmation of their
classification. 

\begin{deluxetable*}{lllccc}
\tabletypesize{\scriptsize}
\tablecaption{NLTT White Dwarfs\label{tbl1}} 
\tablewidth{0pt}
\tablehead{
\colhead{NLTT} & \colhead{Alternate Names} & \colhead{$\mu$,$\theta$\tablenotemark{a}} & \colhead{$V$} & \colhead{$V-J$\tablenotemark{b}} & \colhead{$J-H$\tablenotemark{b}} \\
\colhead{} & \colhead{} & \colhead{($\arcsec$ yr$^{-1}$,deg)} & \colhead{(mag)} & \colhead{(mag)} & \colhead{(mag)} \\
}
\startdata
  287 & LP 464-57 & 0.424, 124.0  & $16.02\pm0.2$\tablenotemark{a} & $0.91\pm0.50$ & $0.026\pm0.100$ \\
  529 & GD 5, LP 192-41 & 0.237, 193.4  & $15.23\pm0.2$\tablenotemark{a} & $0.68\pm0.37$ & $0.214\pm0.067$ \\
 3022 & G 240-93, PHL 3101 & 0.210, 268.9  & $16.67\pm0.2$\tablenotemark{a} & $0.02\pm3.22$ & $-0.362\pm1.027$ \\
 3915 & LP 294-61 & 0.227, 219.3  & $16.15\pm0.2$\tablenotemark{a} & $0.91\pm0.54$ & $0.211\pm0.128$ \\
 6275 & G 72-40, G94-21 & 0.213,  76.6  & $15.77\pm0.2$\tablenotemark{a} & $0.67\pm0.56$ & $0.035\pm0.121$ \\
 8435 & LHS 1421 & 0.611, 190.0  & $15.74\pm0.07$\tablenotemark{c}& $1.28\pm0.27$ & $0.144\pm0.081$ \\
 8581 & G 36-29 & 0.340, 117.0  & $16.29\pm0.2$\tablenotemark{a} & $1.37\pm0.40$ & $0.327\pm0.077$ \\
 8651 & GD 32 & 0.172, 235.7  & $16.23\pm0.2$\tablenotemark{a} & $0.73\pm0.50$ & $0.292\pm0.118$ \\
 9933 & LP 299-22 & 0.190, 119.6  & $15.83\pm0.2$\tablenotemark{a} & $-0.76\pm1.07$& $0.035\pm0.335$ \\
10398 & GD 44 & 0.176, 153.7  & $15.74\pm0.2$\tablenotemark{a} & $0.19\pm0.71$ & $0.034\pm0.166$\\
14307 & G 84-26 & 0.293, 142.9  & $15.30\pm0.2$\tablenotemark{a} & $-0.05\pm0.58$& $0.160\pm0.123$ \\
14352 & LP 656-32 & 0.239, 157.6  & $16.80\pm0.2$\tablenotemark{a} & $0.76\pm0.71$ & $0.063\pm0.200$ \\
17285 & LP 661-15 & 0.211, 312.2  & $16.73\pm0.2$\tablenotemark{a} & $0.84\pm0.74$ & $0.513\pm0.186$ \\
18555 & LP 207-50 & 0.420, 165.8  & $17.45\pm0.2$\tablenotemark{a} & $1.70\pm1.63$ & $-0.029\pm0.251$ \\
18642 & LP 207-55, KUV 07531+4148 & 0.322, 181.4  & $16.56\pm0.2$\tablenotemark{a} & $0.56\pm0.83$ & $0.177\pm0.228$ \\
19019 & LP 311-2, GD 262 & 0.182, 113.3  & $16.09\pm0.2$\tablenotemark{a} & $0.65\pm0.55$ & $0.065\pm0.126$ \\
19138 & G 111-64 & 0.258, 166.5  & $15.74\pm0.2$\tablenotemark{a} & $1.41\pm0.37$ & $0.213\pm0.068$ \\
20165 & KUV 08422+3813 & 0.217, 152.4  & $16.03\pm0.2$\tablenotemark{a} & $0.52\pm0.62$ & $0.081\pm0.138$ \\
21241 & LP 210-58 & 0.200, 258.6  & $17.27\pm0.2$\tablenotemark{a} & $0.87\pm1.66$ & $0.310\pm0.276$ \\
27901 & GD 311, LP 94-285, LB 2052 & 0.232, 238.7  & $16.66\pm0.2$\tablenotemark{a} & $0.62\pm0.80$ & $0.156\pm0.212$ \\
28772 & LP 129-587, G197-35 & 0.300, 266.8  & $16.67\pm0.2$\tablenotemark{a} & $0.15\pm1.73$ & $0.210\pm0.306$ \\
28920 & LP 19-411 & 0.282, 209.3  & $15.90\pm0.2$\tablenotemark{a} & $0.20\pm0.91$ & $0.062\pm0.237$ \\
29233 & G122-61, CBS 451 & 0.300, 262.8  & $16.00\pm0.2$\tablenotemark{a} & $0.77\pm0.52$ & $0.044\pm0.103$ \\
30738 & LP 435-109 & 0.162, 260.8  & $16.19\pm0.2$\tablenotemark{a} & $0.37\pm0.75$ & $0.384\pm0.189$ \\
31347 & LP217-47, LHS 5222 & 0.506, 315.0  & $17.53\pm0.2$\tablenotemark{a} & $0.96\pm1.65$ & $0.084\pm0.259$ \\
31748 & PG 1242-106 & 0.349, 257.1  & $14.43\pm0.2$\tablenotemark{a} & $0.19\pm0.35$ & $0.194\pm0.057$ \\
32390 & LP 96-9 & 0.233, 243.3  & $17.57\pm0.2$\tablenotemark{a} & $1.42\pm0.62$ & $0.673\pm0.188$ \\
32695 & LP 616-70 & 0.190, 271.9  & $16.84\pm0.2$\tablenotemark{a} & $0.40\pm3.22$ & $0.408\pm1.026$ \\
34264 & LP 617-70 & 0.230, 175.5  & $16.46\pm0.2$\tablenotemark{a} & $0.03\pm1.59$ & $0.247\pm0.233$ \\
34564 & LP 678-8 & 0.245, 280.3  & $16.40\pm0.2$\tablenotemark{a} & $0.77\pm0.68$ & $0.155\pm0.159$ \\
35880 & LP 97-430 & 0.309, 281.0  & $17.63\pm0.2$\tablenotemark{a} & $0.91\pm1.63$ & $1.121\pm0.282$ \\
38499 & LP 801-14, EC 14473-1901 & 0.269, 288.0  & $15.80\pm0.02$\tablenotemark{d}& $0.73\pm0.29$ & $0.215\pm0.094$ \\
40489 & LP 503-7, GD 184 & 0.179, 177.2  & $16.56\pm0.2$\tablenotemark{a} & $0.43\pm1.58$ & $0.226\pm0.228$ \\
40607 & G 137-24 & 0.248, 221.6  & $15.07\pm0.05$\tablenotemark{e}& $0.12\pm0.25$ & $0.237\pm0.076$ \\
40881 & LP 273-64, LTT 14655, GD 187 & 0.186, 295.4  & $15.03\pm0.2$\tablenotemark{a} & $0.24\pm0.40$ & $0.107\pm0.084$ \\
40992 & LP 384-38 & 0.417, 173.2  & $16.95\pm0.2$\tablenotemark{a} & $1.19\pm0.77$ & $0.302\pm0.178$ \\
41800 & G 202-26 & 0.318, 330.3  & $16.49\pm0.2$\tablenotemark{a} & $0.81\pm0.69$ & $0.113\pm0.168$ \\
42050 & LP 274-53 & 0.255, 172.7  & $16.95\pm0.2$\tablenotemark{a} & $1.06\pm0.71$ & $0.170\pm0.182$ \\
42153 & LP 861-31 & 0.194, 312.4  & $15.46\pm0.2$\tablenotemark{a} & $0.35\pm0.56$ & $0.045\pm0.126$ \\
43806 & LP 276-33 & 0.329, 177.6  & $15.86\pm0.2$\tablenotemark{a} & $0.31\pm0.61$ & $0.211\pm0.154$ \\
43827 & LP 387-21 & 0.208, 164.7  & $16.90\pm0.2$\tablenotemark{a} & $0.77\pm0.79$ & $0.552\pm0.220$ \\
43985 & LP 331-27 & 0.343, 224.8  & $16.92\pm0.2$\tablenotemark{a} & $0.42\pm2.45$ &$-0.091\pm1.040$ \\
44000 & G 203-39 & 0.385, 228.9  & $16.72\pm0.2$\tablenotemark{a} & $1.28\pm0.60$ & $0.266\pm0.144$ \\
44149 & LP 276-48 & 0.250, 135.3  & $17.98\pm0.2$\tablenotemark{a} & $1.41\pm3.24$ & $0.231\pm1.039$ \\
44447 & LP 226-48 & 0.195, 355.4  & $16.13\pm0.2$\tablenotemark{a} & $0.49\pm0.61$ & $0.276\pm0.141$ \\
44986 & EG 545, GD 362, PG 1729+371 & 0.224, 173.1  & $16.23\pm0.2$\tablenotemark{a} & $0.05\pm3.20$ & $0.128\pm1.005$ \\
45344 & LP 227-31 & 0.217, 189.1  & $17.25\pm0.2$\tablenotemark{a} & $1.43\pm1.53$ & $0.158\pm0.191$ \\
45723 & LP 228-12 & 0.334,   2.9  & $17.26\pm0.2$\tablenotemark{a} & $1.32\pm0.76$ & $0.081\pm0.202$ \\
49985 & LP 872-20 & 0.278, 130.0  & $15.42\pm0.2$\tablenotemark{a} & $0.51\pm0.40$ & $0.224\pm0.076$ \\
50029 & LP 872-48, BPS CS 22880-0126 & 0.136,   3.1  & $15.01\pm0.06$\tablenotemark{c}& $0.41\pm0.28$ & $0.012\pm0.087$ \\
51103 & LP 637-32 & 0.291, 209.2  & $16.59\pm0.2$\tablenotemark{a} & $0.91\pm0.80$ & $0.042\pm0.233$ \\
52404 & LP 930-61 & 0.234,  44.0  & $16.26\pm0.17$\tablenotemark{c}& $1.05\pm0.46$ & $0.115\pm0.104$ \\
53177 & LP 759-50, HE 2209-1444 & 0.267,  70.4  & $15.09\pm0.02$\tablenotemark{f}& $0.59\pm0.23$ & $0.134\pm0.084$ \\
53447 & LP 287-39 & 0.463,  79.3  & $16.99\pm0.2$\tablenotemark{a} & $1.57\pm0.57$ & $0.236\pm0.136$ \\
53468 & LP 619-64, PHL 218 & 0.249,  72.0  & $15.3\pm0.2$\tablenotemark{a}  & $-0.57\pm0.90$ &$-0.025\pm0.243$ \\
53996 & LP 400-6 & 0.320, 134.9  & $17.95\pm0.2$\tablenotemark{a} & $1.59\pm1.58$  &$0.347\pm0.235$ \\
54047 & LHS 3821, LP 520-28 & 0.538,  68.4  & $17.82\pm0.2$\tablenotemark{a} & $1.47\pm1.57$  &$0.348\pm0.228$ \\
55932 & G 275-8, LP 877-69, LTT 9373 & 0.339, 106.0  & $13.68\pm0.06$\tablenotemark{c}& $-0.49\pm0.26$ & $0.003\pm0.075$ \\
56122 & LP 522-17 & 0.294, 206.3  & $17.29\pm0.2$\tablenotemark{a} & $1.39\pm0.62$ & $0.451\pm0.155$ \\
56805 & LP 522-46 & 0.352,  71.0  & $15.81\pm0.2$\tablenotemark{a} & $1.28\pm0.42$ & $0.173\pm0.077$ \\
58283 & LP 463-68 & 0.225, 148.3  & $16.57\pm0.2$\tablenotemark{a} & $0.63\pm0.71$ & $0.108\pm0.191$ \\
\enddata
\tablenotetext{a}{From \citet{sal2003}.}
\tablenotetext{b}{$JHK$ from 2MASS magnitudes converted to CIT.}
\tablenotetext{c}{From SPM Catalog 2.0.}
\tablenotetext{d}{From \citet{kil1997}.}
\tablenotetext{e}{From \citet{egg1968}.}
\tablenotetext{f}{From \citet{bee1992}.}
\end{deluxetable*}{}

\subsection{Spectroscopy}

The rNLTT white dwarf candidates were observed using the Dual Imaging 
Spectrogram (DIS) 
attached to the 3.5 m telescope at the Apache Point Observatory (APO) on
2004 May 31, June 1, 2005 May 12, 16, June 2, July 2, 15, August 9 and
November 12, 14. 
We used 300 lines/mm gratings to obtain a spectral range of 3700 to 5600 \AA\
with a dispersion of 2.4 \AA\ per pixel in the blue, and a spectral range of 
5300 to 9800 \AA\ with a dispersion of 2.3 \AA\ in the red.

We also obtained spectroscopy by cross-correlating 
the list of rNLTT white dwarf candidates with the Sloan Digital Sky Survey 
(SDSS). We used the archival spectra from the 4th Data Release of SDSS in our
analyses of these objects.

\begin{figure*}
\plotone{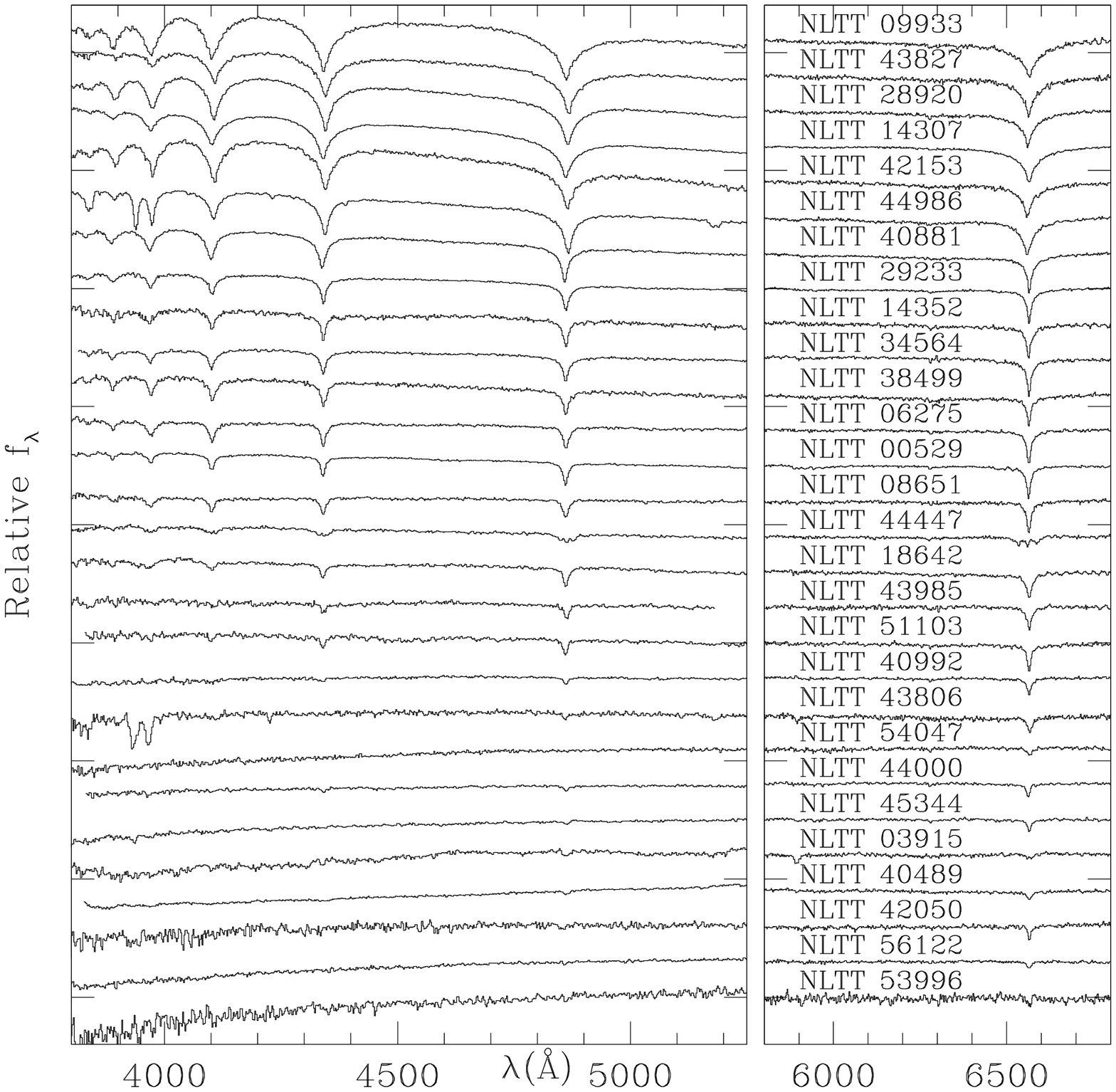}
\caption{APO spectra of NLTT hydrogen-rich white dwarfs. \label{fig1}}
\end{figure*}

\begin{figure}
\plotone{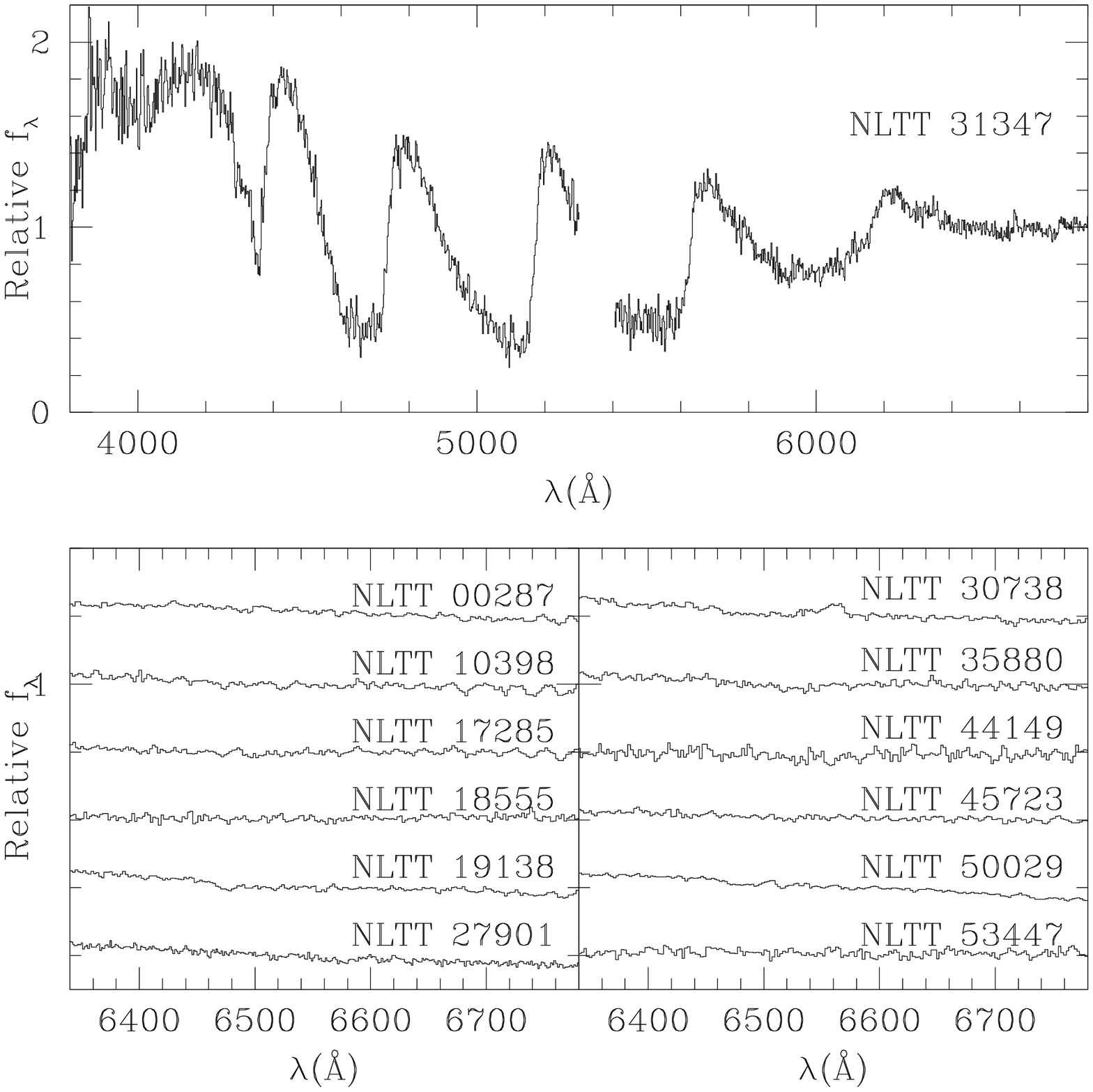}
\caption{Spectra of NLTT helium-rich white dwarfs. {\it Top:} Spectrum
of the DQ white dwarf NLTT31347. {\it Bottom:} Spectra of the nine new DC
white dwarfs. Note the possible weak H$\alpha$ emission in NLTT 30738.
\label{fig2}}
\end{figure}

Figure~\ref{fig1} shows the 
spectra of all the hydrogen-rich white dwarfs, and Figure~\ref{fig2} shows the
spectra of the helium-rich white dwarfs obtained at APO.

\subsection{Photometry}

Table~\ref{tbl1} lists the $V$ magnitudes and $JHK$ photometry for the
rNLTT white dwarf candidates for which we have obtained spectroscopy.
The table also lists the stars' alternate names and their proper motion taken
from rNLTT catalog of \citet{sal2003}.
The $V$ magnitudes are taken from the rNLTT catalog of \citet{sal2003}, except for
stars where more accurate photometry from another source was available. The
$JHK$ photometry was obtained from the 2MASS database available at the
Centre de Donn\'ees Astronomique de Strasbourg. The data were converted from 
2MASS system to the CIT system using color transformations provided by
\citet{cut2003}. \footnote{Available on the World Wide Web at 
http://www.ipac.caltech.edu/2mass/releases/allsky/doc/explsup.html.}

We cross-correlated the list of rNLTT white dwarf candidates with the SDSS to
obtain {\it ugriz} photometry (Table~\ref{tbl3}).

\begin{deluxetable*}{llccc}
\tabletypesize{\scriptsize}
\tablecaption{SDSS colors of NLTT White Dwarfs \label{tbl3}}
\tablewidth{0pt}
\tablehead{
\colhead{NLTT} & \colhead{SDSS} & \colhead{$u-g$} & \colhead{$g-r$} & \colhead{$r-i$} \\
\colhead{} & \colhead{} & \colhead{(mag)} & \colhead{(mag)} & \colhead{(mag)} \\
}
\startdata
 3022 & J005438.84-095219.8 & $0.404\pm0.013$ &  $0.055\pm0.008$ & $-0.071\pm0.009$ \\
18555 & J075313.27+423001.5 & $1.929\pm0.049$ &  $0.885\pm0.011$ &  $0.319\pm0.010$ \\
18642 & J075631.11+413950.9 & $0.597\pm0.013$ &  $0.193\pm0.008$ &  $0.073\pm0.009$ \\
19019 & J080946.19+292032.1 & $0.533\pm0.010$ &  $0.170\pm0.007$ &  $0.046\pm0.008$ \\
19138 & J081411.16+484529.6 & $0.562\pm0.008$ &  $0.245\pm0.006$ &  $0.020\pm0.007$ \\
20165 & J084524.62+380156.0 & $0.467\pm0.010$ &  $0.045\pm0.009$ & $-0.020\pm0.010$ \\
21241 & J091356.83+404734.6 & $0.561\pm0.016$ &  $0.208\pm0.010$ &  $0.047\pm0.011$ \\
27901 & J113534.61+572451.7 & $0.326\pm0.011$ &  $0.086\pm0.007$ & $-0.031\pm0.008$ \\
28772 & J115123.93+541147.7 & $0.445\pm0.013$ & $-0.075\pm0.009$ & $-0.105\pm0.010$ \\
29233 & J120003.28+433541.5 & $0.469\pm0.010$ &  $0.071\pm0.008$ & $-0.013\pm0.008$ \\
31347 & J123752.12+415625.8 & $0.105\pm0.018$ &  $0.640\pm0.010$ &  $0.257\pm0.010$ \\
32390 & J125629.36+610200.9 & $0.455\pm0.016$ & $-0.036\pm0.011$ & $-0.102\pm0.013$ \\
32695 & J130247.97-005002.7 & $0.454\pm0.013$ & $-0.136\pm0.011$ & $-0.142\pm0.013$ \\
34264 & J132937.15-013430.5 & $0.526\pm0.017$ &  $0.148\pm0.010$ &  $0.040\pm0.010$ \\
35880 & J135758.43+602855.2 & $0.776\pm0.028$ &  $0.351\pm0.012$ &  $0.106\pm0.013$ \\
40881 & J154033.31+330852.8 & $0.402\pm0.008$ &  $0.000\pm0.006$ & $-0.077\pm0.007$ \\
40992 & J154234.65+232939.8 & $0.951\pm0.017$ &  $0.391\pm0.010$ &  $0.149\pm0.010$ \\
41800 & J160112.70+531700.0 & $0.613\pm0.013$ &  $0.201\pm0.008$ &  $0.055\pm0.009$ \\
42050 & J160714.18+342345.7 & $1.116\pm0.025$ &  $0.517\pm0.010$ &  $0.183\pm0.010$ \\
43806 & J165445.69+382936.5 & $0.993\pm0.017$ &  $0.417\pm0.010$ &  $0.170\pm0.010$ \\
43827 & J165538.92+253345.9 & $0.225\pm0.014$ & $-0.106\pm0.010$ & $-0.159\pm0.011$ \\
43985 & J165939.99+320320.0 & $0.689\pm0.022$ &  $0.291\pm0.012$ &  $0.087\pm0.012$ \\
44149 & J170447.70+360847.4 & $1.832\pm0.062$ &  $0.785\pm0.014$ &  $0.308\pm0.013$ \\
56122 & J231206.08+131057.6 & $1.499\pm0.030$ &  $0.635\pm0.011$ &  $0.225\pm0.010$ \\
56805 & J232519.88+140339.7 & $1.592\pm0.018$ &  $0.598\pm0.009$ &  $0.287\pm0.010$ \\
\enddata
\end{deluxetable*}

\section{Model Atmospheres and Spectra}

The stars presented in this paper have been analyzed for their $T_{\rm eff}$,
$\log{g}$ using a grid of computed pure hydrogen LTE plane-parallel models.
The grid of models extend from 
$T_{\rm eff} = 4500$ to 6500 K (in steps of 500 K) at $\log{g} = 7.0, 8.0$ and 
$9.0$, from $T_{\rm eff} = 7000$ to 16000 K (in steps of 1000 K), from 18000 to 
32000 K (in steps of 2000 K) and from 36000 to 84000 K (in steps of 4000 K) at 
$\log{g} = 7.0$ to $9.5$ (in steps of 0.25 dex).
Convective energy transport in cooler atmospheres is included by applying the
Schwartzschild stability criterion and by using the mixing length formalism
described by \citet{mih1978}. We have assumed the ML2 parameterization of the
convective flux \citep{fon1981} and adopting $\alpha = 0.6$ \citep{ber1992}, 
where $\alpha$ is the ratio of the mixing length to the pressure scale height.
The equation of convective energy transfer was
fully linearized within the Feautrier solution scheme and subjected to the 
constraint that $\mathcal{F}_{\rm total} = \sigma_R T^4_{\rm eff} =
\mathcal{F}_{\rm conv}+\mathcal{F}_{\rm rad}$, where $\mathcal{F}_{\rm conv}$
is the convective flux and $\mathcal{F}_{\rm rad}$ is the radiative flux.

The dissolution of the hydrogen energy levels in the high-density atmospheres
of white dwarfs was calculated using the formalism of 
\citet[][hereafter HM]{hum1988} and following the treatment of 
\citet[][hereafter HHL]{hub1994}).  For the sake of reproducibility
we now provide details of the calculations as included in our fortran code.
That is we calculated the occupation
probability from charged particles of level $n$ using the form 
(HM Eq. 4.26 and HHL Eq. A.2):
\begin{equation}
w_n ({\rm charged}) = Q(\beta_n)=\int^\beta_0 W(\beta) d\beta. 
\end{equation}
$W(\beta)$ is the Holtzmark distribution function. $\beta$ can be calculated using:
\begin{equation}
\beta = K_n\Big(\frac{Z^3}{16n^4}\Big)\Big(\frac{4\pi a_0}{3}\Big)^{-\frac{2}{3}}N_e^{-1}N_{\rm ion}^{\frac{1}{3}},
\end{equation}
where $Z$ is the ionic charge, $N_e$ is the electron number density 
$N_{\rm ion}$ is the ion number density and $a_0$ is Bohr radius 
($a_0 = 0.529\times10^{-8}$ cm). And assuming that $N_e = N_{\rm ion}$ we can
simplify $\beta$ to:
\begin{equation}
\beta = 8.59\times10^{14} \Big(\frac{K_n Z^3 N_e^{-\frac{2}{3}}}{n^4}\Big).
\end{equation}
$K_n$ is defined by:
\begin{equation}
K_n = \left\{ \begin{array}{ll}
      1 & {\rm for}\ n \leq 3\\
      \frac{16}{3} \Big( \frac{n}{n+1}\Big)^2 \Big(\frac{n+7/6}{n^2+n+1/2}\Big) & {\rm for}\ n > 3 \\
      \end{array} \right.
\end{equation}
We can approximate $Q(\beta_n)$ by the rational expression presented by HHL:
\begin{equation}
w_n ({\rm charged}) = \frac{f}{1+f},
\end{equation}
where $f$ is given by:
\begin{equation}
f=\frac{0.1402(x+4Za^3)\beta^3}{1+0.1285x\beta^\frac{3}{2}},
\end{equation}
and $x=(1+a)^{3.15}$ and $a$ is the correlation parameter which is defined as
\begin{equation}
a=f^\prime 0.09\sqrt{\frac{2}{T}} \Big(\frac{N_e^\frac{1}{6}}{1+\frac{N_{HI}}{N_e}}\Big).
\end{equation}
Here, $T$ is the temperature and $N_{HI}$ is the number density of neutral 
hydrogen. We introduced the scaling factor $f^\prime$ to the correlation parameter 
$a$ which can have values between 0 and 1. In calculating our model spectra
we set $f^\prime = 1$. Lowering the value for $f^\prime$ will generate lower values in
surface gravity measurements.

HM also considered the contribution to the occupation probability from neutral 
particles, which is necessary in the case of cool white dwarfs where neutral
hydrogen becomes dominant. Therefore, the neutral particle contribution to the 
occupation probability is given by:
\begin{equation}
w_n=exp\Big(-\frac{4\pi}{3} \sum_m N_{m}(r_n+r_m)^3 \Big)
\end{equation}
where we can assume that the effective interaction radius is a fraction $f$ of
the hydrogen atomic radius which is associated with level $n$ i.e., 
$r_n = fn^2 a_0$. For our calculations we have assumed $f=0.5$ \citep[see][]{ber1991}, and therefore
\begin{equation}
w_n=exp\Big(\frac{-\pi a_0^3}{6} \sum_m N_{m}(n^2+m^2)^3 \Big).
\end{equation}
And finally, the combined occupation probability is the product of the
occupation probabilities from neutral and charged particles, i.e.,
\begin{equation}
w_{n, {\rm comb}} = w_n ({\rm charged}) \times w_n ({\rm neutral})
\end{equation}

Since the occupation probability depends on the number density of neutral
hydrogen $N_{HI}$, we first obtain an estimate of the populations for all
levels by including the contribution from only neutral particles, i.e., $m=1$.
We then repeat the calculations by including the contribution from both
neutral and charged particles, until the $w_n$ remains constant for all $n$.

The calculated level occupation
probabilities are then explicitly included in the calculation of the line and
continuum opacities. The Balmer line profiles are calculated using the tables
of Stark-broadened \ion{H}{1} line profiles of \citet{lem1997} convolved with  
normalized resonance line profiles. We have adopted the Stark-broadened line 
profiles of \citet{lem1997} over those of \citet{sch1994}
because the tables of \citet{lem1997} are complete at higher electron
densities ($10^{10} \le n_e \le 10^{18}$ cm$^{-3}$). The tables of 
\citet{sch1994} are incomplete at higher electron densities for many lines.

We modeled the heavy element lines using Voigt profiles \citep{gra1992} including
Stark and van der Waals broadening parameters.
In cool DA white dwarfs which display metal lines van der Waals
broadening is dominant and the pertubers are neutral hydrogen atoms. For
atmospheres dominated by hydrogen, we have used an approximation of the damping
constant, $\gamma$ as defined by \citet{gra1992}:
\begin{equation}
\log{\gamma_6} = 19.6 +0.4\log{C_6(H)}+\log{P_g}-0.7\log{T}
\end{equation}
where $P_g$ is the pressure, T is the temperature and $C_6(H)$ can be calculated
using:
\begin{equation}
C_6(H) = 0.3\times10^{-30}\Big(\frac{1}{(I-\chi-\chi_\lambda)^2}-\frac{1}{(I-\chi)^2}\Big).
\end{equation}
Here, $I$ is the ionization potential, $\chi$ is the excitation potential of
the lower level for the heavy metal atom. And $\chi_\lambda$ is the energy
of a photon in the line, that is $\chi_\lambda = 1.24\times10^4/\lambda$, where
$\lambda$ is in \AA.
 
\section{Analysis and Results}

\subsection{Balmer lines}

The Balmer lines of the hydrogen-rich white dwarfs were analyzed using a
$\chi^2$ minimization technique and our cool hydrogen model atmospheres. The quoted
uncertainties are only statistical ($1\sigma$) and do not take into account
possible systematic effects in model calculations or data acquisition and
reduction procedures.

\subsection{Sloan Colors}

Using the computed spectral grid, we have computed SDSS synthetic colors. For 
the white dwarf candidates for which we obtained spectra and for which 
SDSS photometry is available, we have compared the observed colors to the theoretical colors
as a check of the temperatures obtained from the Balmer line profile fits.
We have also calculated SDSS colors for a black-body spectrum for temperatures
ranging from 4000 to 36000 K (in 1000 K intervals), and used these colors
to obtain temperature estimates of DC white dwarfs.

\begin{figure}
\plotone{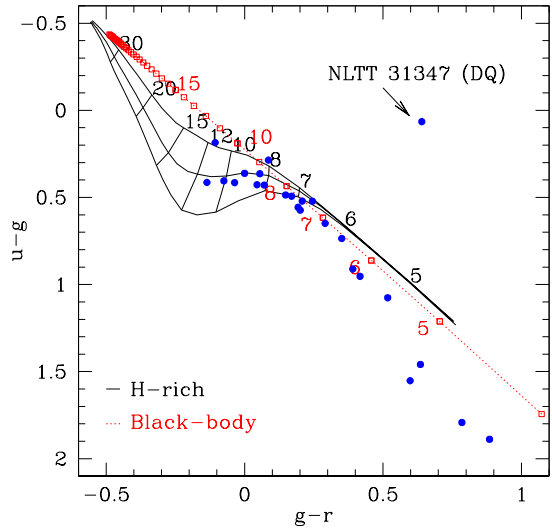}
\caption{SDSS $u-g$ versus $g-r$ photometry of NLTT white dwarfs compared
to synthetic colors of H-rich white dwarfs ({\it full line}) and blackbody 
({\it dotted line}) colors. The DQ white dwarf NLTT 31347 is marked. The
effective temperature is indicated in units of 1000 K and for H-rich colors
$\log{g} = 7.0$, 8.0 and 9.0 from bottom to top.
\label{fig_gmr_umg_ext}}
\end{figure}

\begin{figure}
\plotone{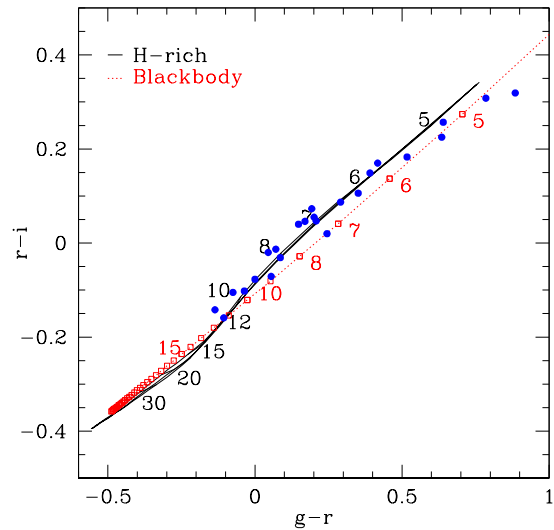}
\caption{SDSS $r-i$ versus $g-r$ photometry of NLTT white dwarfs compared
to synthetic colors of H-rich white dwarfs ({\it full line}) and blackbody
({\it dotted line}) colors. The
effective temperature is indicated in units of 1000 K.
\label{fig_gmr_rmi_ext}}
\end{figure}

The observed colors for a number of white dwarfs are compared to the 
synthetic $u-g$ versus $g-r$ colors in Figure~\ref{fig_gmr_umg_ext} and
$r-i$ versus $g-r$ colors in Figure~\ref{fig_gmr_rmi_ext}. The $u-g$/$g-r$ 
diagram shows that the observed $u-g$ is larger than the $u-g$ calculated 
from the model atmospheres for H-rich white dwarfs with temperatures less than
$T_{\rm eff} = 7000$ K \citep[also noted by][]{kil2006}, hence implying some missing opacity in the $u$ band.
\citet{ber2001} suggests that a similar effect in the $B$ band may be
due to the bound-free opacity associated with the dissolved 
atomic levels of hydrogen, i.e., that following a 
bound-bound transition there
is a probability that the upper level may be sufficiently perturbed by the
surrounding particles that the electron will no longer be bound to the nucleus.

\begin{figure}
\plotone{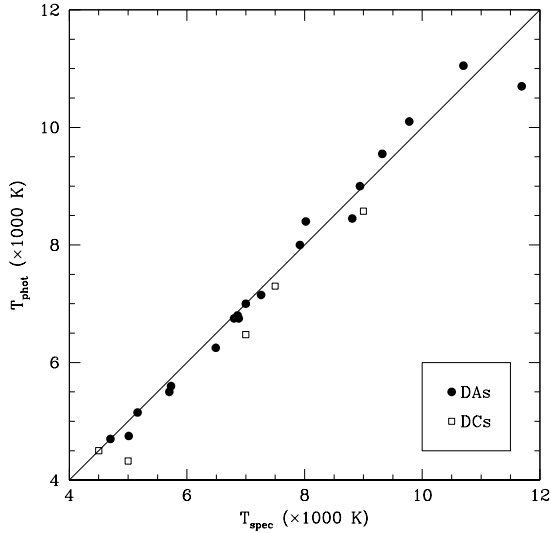}
\caption{A comparison of the effective temperatures obtained using the spectra
of stars (DA, {\it full circles}; DC, {\it open squares}) to the temperatures obtained using the $u-g$/$g-r$ and $r-i$/$g-r$
colors. \label{fig_temp_comp}}
\end{figure}

For the stars with SDSS $ugriz$ photometry, the temperatures we obtained from 
spectroscopic fits were compared to the temperatures obtained using the
$u-g/g-r$ and $r-i/g-r$ colors (Fig.~\ref{fig_temp_comp}). 
For DA white dwarfs, the label ``spectroscopic fits'' refers to temperatures obtained by
fitting the Balmer line profiles with model spectra, and for DC white dwarfs the same
label refers to temperatures obtained by comparing a black-body
to the observed spectrum.
The only
point that appears to be in disagreement (i.e., the photometric temperature
is lower than the spectroscopic temperature), is the ultramassive white
dwarf NLTT 43827, which has a spectroscopic temperature of 11690 K. At this
temperature the effect of gravity on the $u-g/g-r$ colors is the strongest and
since we assumed $\log{g} = 8$ in our photometric temperature estimates, it
is likely to effect the final temperature estimate for white dwarfs with
significantly different surface gravities in this region of the color diagram.
In Figure~\ref{fig_temp_comp} the temperatures determined from the photometry 
appear to be slightly higher than those
determined from spectroscopy for $T_{\rm eff} > 8000$ K and vice versa for
$T_{\rm eff} < 8000$ K. This may be due to measurement errors or possibly
to systematic errors in either spectroscopic (i.e., via the fitting of Balmer lines
with model spectra) or photometric calibration.

\subsection{Interesting White Dwarfs}

Table~\ref{tbl2} summarizes the results of the analyses of the white dwarfs.
The table also includes white dwarfs from \citet{ven2003} and
\citet{kaw2004}. The temperatures and surface gravities of the DA white dwarfs 
from these papers were remeasured with the improved synthetic spectra. For
the cool DA white dwarfs the gravities went down significantly, in particular
the gravity ($\log{g} = 9.00$) of NLTT 49985 went down to $\log{g} = 8.31$
and the high gravity should be discarded. Also the surface gravity of
NLTT 31748 went down from $\log{g} = 8.40$ to $\log{g} = 7.85$.

\begin{deluxetable*}{lllccccccr}
\tabletypesize{\scriptsize}
\tablecaption{White Dwarf Parameters and Kinematics. \label{tbl2}}
\tablewidth{0pt}
\tablehead{
\colhead{NLTT} & \colhead{} & \colhead{WD} & \colhead{Spectral type} & \colhead{$T_{\rm eff}$} & \colhead{$\log{g}$} & \colhead{M} & \colhead{$M_V$} & \colhead{$d$} & \colhead{$U,V,W$} \\
\colhead{} & \colhead{} & \colhead{} & \colhead{} & \colhead{(K)} & \colhead{(cgs)} & \colhead{($M_\odot$)} & \colhead{(mag)} & \colhead{(pc)} & \colhead{(km s$^{-1}$)}\\
}
\startdata
  287 & APO  & 0004$+$122 & DC  & $6300^{+1500}_{-1100}$ & (8.0) & (0.57)       & 13.91 & 26 & -14, -33, -19 \\
  529 & APO  & 0008$+$423\tablenotemark{a} & DA  &  $7380\pm60$ & $8.38\pm0.08$  & $0.83\pm0.05$ & 13.96 & 18 & 20, 3, -10 \\
 3022 & SDSS & 0052$-$101 & DA  &  $8810\pm50$ & $8.11\pm0.06$  & $0.66\pm0.04$ & 12.86 & 58 & 59, 36, 6 \\
 3915 & APO  & 0108$+$277 & DAZ & $5270\pm250$ & $8.36\pm0.60$  & $0.81\pm0.37$ & 15.41 & 14 & 21, 5, -3, \\
 6275 & APO  & 0150$+$256 & DA  & $7530\pm70$  & $8.18\pm0.11$  & $0.70\pm0.07$ & 13.56 & 28 & -12, -8, 19 \\
 8435 & MSO/SSO & 0233$-$242\tablenotemark{a,b} & DC & $5400\pm500$&  (8.0)  &   (0.58) & 14.65 & 17 & 49, -24, -1  \\
 8581 & MSO  & 0236$+$259\tablenotemark{b} & DA  & $5500\pm500$ & (8.0)          &   (0.58)         & 14.67 & 21 & -7, -24, 8 \\
 8651 & APO  & 0237$+$315 & DA  & $7040\pm90$  & $8.36\pm0.16$  & $0.82\pm0.10$    & 14.12 & 26 & 22, 10, -10 \\
 9933 & APO  & 0304$+$317 & DA  & $13300\pm300$& $7.87\pm0.10$  & $0.53\pm0.05$    & 11.30 & 81 & -23, -60, 10 \\
10398 & APO  & 0313$+$393 & DC  &$11000\pm1000$& (8.0)          & (0.58)           & 12.02 & 56 & -4, -33, -16 \\
14307 & APO  & 0457$-$004 & DA  & $10800\pm80$ & $9.15\pm0.06$  & $1.24\pm0.02$    & 14.09 & 17 & 17, -17, 10 \\
14352 & APO  & 0458$-$064 & DA  & $7900\pm100$ & $8.65\pm0.13$  & $1.01\pm0.08$    & 14.19 & 33 & 28, -28, 4 \\
17285 & APO  & 0658$-$075 & DC  & $5900\pm1000$& (8.0)          & (0.57)           & 14.20 & 32 & -9, 28, -4 \\
18555 & APO  & 0749$+$426 & DC  & $4300\pm300$ & (8.0)          & (0.57)           & 16.20 & 18 & 9, -31, 8 \\
18642 & APO  & 0753$+$417 & DA  & $6880\pm70$  & $8.62\pm0.16$  & $0.99\pm0.10$    & 14.68 & 24 & 5, -30, 0 \\
19019 & SDSS & 0806$+$294 & DA  &  $7000\pm70$ & $8.12\pm0.11$  & $0.66\pm0.07$ & 13.78 & 29 & 23, -8, 24 \\
19138 & APO  & 0810$+$489\tablenotemark{a} & DC  & $7300\pm100$ & (8.0)          & (0.57)        & 13.31 & 23 & 8, -23, 11 \\
20165 & SDSS & 0842$+$382 & DA  &  $8020\pm50$ & $8.16\pm0.06$  & $0.69\pm0.04$ & 13.31 & 35 & 20, -28, 17 \\
21241 & SDSS & 0910$+$410 & DA  &  $6860\pm100$& $8.00\pm0.25$  & $0.59\pm0.14$ & 13.72 & 51 & -23, -4, -27 \\
27901 & SDSS & 1132$+$574 & DC  & $8600\pm200$ & (8.0)          & (0.58)        & 12.73 & 61 & -32, -47, 7 \\
28772 & SDSS & 1148$+$544 & DA  &  $9780\pm70$ & $8.14\pm0.06$  & $0.68\pm0.04$ & 12.51 & 68 & -73, -40, -13 \\
28920 & APO  & 1151$+$795 & DA  & $10810\pm100$& $8.23\pm0.07$  & $0.74\pm0.04$ & 12.30 & 52 & -45, -33, -12 \\
29233 & APO  & 1157$+$438 & DA  &  $7920\pm50$ & $8.19\pm0.07$  & $0.71\pm0.04$ & 13.40 & 33 & -28, -21, -1 \\
30738 & APO  & 1223$+$188 & DC  &  $6000\pm700$\tablenotemark{c} & (8.0) & (0.57) & 14.12 & 26 & -5, -7, 5 \\
31347 & APO  & 1235$+$422 & DQ  &  $\sim 6250$ & (8.0) &(0.57) & $\sim 15.4$\tablenotemark{d} & 27 & -52, 19, -7 \\
31748 & SSO  & 1242$-$105\tablenotemark{a,b} & DA  &  $8280\pm80$ & $7.85\pm0.13$  & $0.51\pm0.06$ & 12.72 & 22 & -19, -20, 2 \\
32390 & SDSS & 1254$+$613 & DA  &  $9320\pm70$ & $8.18\pm0.08$  & $0.71\pm0.05$ & 12.75 & 92 & -44, -76, 35 \\
32695 & SDSS & 1300$-$005 & DA  & $10700\pm80$ & $8.22\pm0.05$  & $0.74\pm0.03$ & 12.32 & 80 & -51, -33, 11 \\
34264 & SDSS & 1327$-$013 & DA  &  $7260\pm110$& $8.12\pm0.19$  & $0.67\pm0.11$ & 13.62 & 37 &  31, -23, -13 \\
34564 & APO  & 1333$-$054 & DA  & $7790\pm100$ & $8.54\pm0.19$  & $0.94\pm0.12$ & 14.04 & 30 & -19, -11, 16 \\
35880 & APO  & 1356$+$607 & DC  & $6500\pm100$ & (8.0)          & (0.57)        & 13.77 & 59 & -63, -39, 20 \\
38499 & APO  & 1447$-$190 & DA  &  $7660\pm80$ & $7.81\pm0.15$  & $0.49\pm0.06$ & 12.97 & 37 & -21, -15, 37 \\
40489 & APO  & 1529$+$141 & DA  &  $5250\pm200$& $8.00\pm0.50$  & $0.58\pm0.24$ & 14.89 & 22 & 23, -7, 1 \\
40607 & MDM/SSO & 1532$+$129\tablenotemark{a} & DZ & $\sim 7500$& (8.0)          & (0.57)        & 13.20 & 24 & 14, -22, 12\\
40881 & APO  & 1538$+$333 & DA  &  $8940\pm100$& $8.44\pm0.10$  & $0.88\pm0.06$ & 13.35 & 22 & -5, -2, 18 \\
40992 & APO  & 1540$+$236 & DA  & $5730\pm100$ & $8.20\pm0.35$  & $0.71\pm0.21$ & 14.77 & 27 & 52, -25, -5 \\
41800 & SDSS & 1559$+$534 & DA  &  $6800\pm90$ & $8.20\pm0.20$  & $0.72\pm0.13$ & 14.04 & 31 & -36, 3, 14 \\
42050 & APO  & 1605$+$345 & DA  & $5150\pm350$ & $\sim 7.0$     & (0.18)        & 13.73 & 44 & 57, -18, 0 \\
42153 & APO  & 1607$-$250 & DA  & $10420\pm120$& $8.22\pm0.09$  & $0.74\pm0.06$ & 12.41 & 41 & -2, 6, 43 \\
43806 & APO  & 1653$+$385 & DAZ & $5700\pm240$ & $8.28\pm0.50$  & $0.76\pm0.30$ & 14.91 & 15 & 31, -4, 4 \\
43827 & APO  & 1653$+$256 & DA  & $11690\pm140$& $9.35\pm0.05$  & $1.31\pm0.01$ & 14.29 & 33 & 37, -7, -7 \\
43985 & APO  & 1657$+$321 & DA  & $6490\pm80$  & $8.76\pm0.21$  & $1.07\pm0.13$ & 15.20 & 22 & 28, -22, 23 \\
44000 & APO  & 1658$+$445 & DA  & $5420\pm100$ & $7.68\pm0.30$  & $0.42\pm0.12$ & 14.30 & 31 & 38, -32, 39 \\
44149 & APO  & 1703$+$362 & DC  & $4500\pm400$ & (8.0)          & (0.57)        & 15.83 & 27 & 34, 9, -14 \\
44447 & APO  & 1713$+$393 & DAP &  $7000\pm1000$ & magnetic       &(0.59)         & 13.60 & 32 & -18, 14, 13 \\
44986 & APO  & 1729$+$371 & DAZ &  $9760\pm70$ & $9.19\pm0.08$  & $1.26\pm0.03$ & 14.54 & 22 & 31, -2, 0 \\
45344 & APO  & 1741$+$436 & DA  &  $5340\pm160$& $7.84\pm0.50$  & $0.49\pm0.20$ & 14.59 & 34 & 43, -6, 6 \\
45723 & APO  & 1755$+$408 & DC  & $5400\pm600$ & (8.0)          & (0.57)        & 14.65 & 33 & -39, 21, 16 \\
49985 & SSO  & 2048$-$250\tablenotemark{a} & DA  &  $7630\pm150$& $8.31\pm0.31$  & $0.79\pm0.20$ & 13.72 & 22 & -1, -11, -14 \\
50029 & APO  & 2049$-$222 & DC  & $9300\pm1000$& (8.0)          & (0.58)        & 12.49 & 28 & 5, 22, 11 \\
51103 & APO  & 2119$-$017 & DA  & $6440\pm120$ & $8.08\pm0.27$  & $0.64\pm0.16$ & 14.09 & 32 & 43, -24, 5 \\
52404 & SSO  & 2152$-$280\tablenotemark{a} & DC  & $6500\pm500$ & (8.0)          & (0.57)        & 13.77 & 32 & -15,27,-5  \\
53177 & MSO  & 2209$-$147\tablenotemark{b,e} & DA  & $7920\pm100$ & $7.84\pm0.30$  & $0.50\pm0.12$ & 12.89 & 28 & -21, 12, -7 \\
53447 & APO  & 2215$+$368 & DC  & $4750\pm250$ & (8.0)          & (0.57)        & 15.44 & 20 & -30, -1, -10 \\
53468 & SSO  & 2215$-$204\tablenotemark{a} & DA  &$15120\pm500$ & $7.75\pm0.15$  & $0.48\pm0.05$ & 10.89 & 76 & -69,15,-35  \\
53996 & APO  & 2227$+$232 & DA  & $5000\pm500$ & $\sim 7.0$     & (0.18)        & 13.92 & 64 & -9, -40, -77 \\
54047 & APO  & 2228$+$151 & DA  & $5580\pm260$ & $8.68\pm0.58$  & $1.02\pm0.37$ & 15.70 & 27 & -57, 5, -9  \\
55932 & SSO  & 2306$-$220\tablenotemark{a} & DA  & $14810\pm260$& $7.86\pm0.08$  & $0.53\pm0.04$ & 11.08 & 33 & -28, -25, -14 \\
56122 & APO  & 2309$+$129 & DA  &  $5010\pm200$& $7.48\pm0.60$  & $0.33\pm0.18$ & 14.45 & 37 & 53, -15, -14 \\
56805 & MSO  & 2322$+$137\tablenotemark{b} & DA  & $4700\pm300$ & $\sim 7.0$ & (0.18) & 14.27 & 20 & -23 0, 4 \\
58283 & APO  & 2350$+$205 & DA  &  $7380\pm50$ & $7.95\pm0.10$  & $0.56\pm0.06$ & 13.31 & 45 & 6, -27, -28 \\
\enddata
\tablenotetext{a}{Also in \citet{kaw2004}.}
\tablenotetext{b}{Also in \citet{ven2003}.}
\tablenotetext{c}{Possible binary, $T_{eff}$ estimate from $V-J/J-H$ diagram.}
\tablenotetext{d}{See text.}
\tablenotetext{e}{Double degenerate system.}
\end{deluxetable*}

\begin{figure*}
\plotone{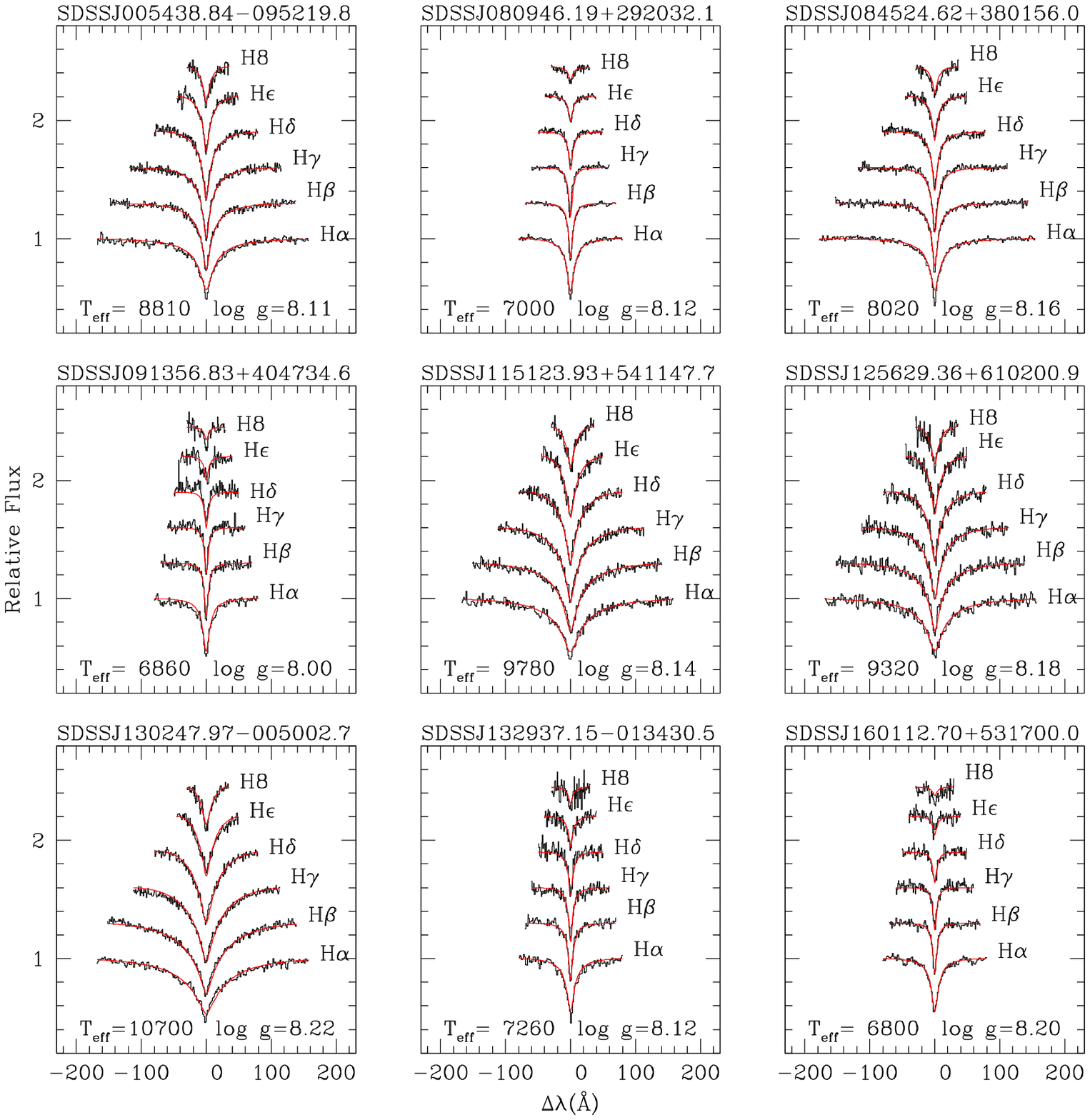}
\caption{Balmer line profiles of NLTT white dwarfs with SDSS spectra compared
to synthetic spectra. \label{fig_SDSS_balmer}}
\end{figure*}

\paragraph{NLTT 529} was observed by \citet{kaw2004} and they classified this 
object as a cool DA white dwarf. We have reobserved this star and fitted
the Balmer line profiles (H$\alpha$ - H9) to obtain an effective temperature
of $7380\pm60$ K and $\log{g} = 8.38\pm0.08$. 
\paragraph{NLTT 3022} is also known as G 240-93 and PHL 3101, and was  
observed as part of the SDSS as 
SDSS J005438.3-095219.8. \citet{kle2004} obtained $T_{\rm eff} = 8900$ K
and $\log{g} = 8.20$. We fitted the Balmer lines with model spectra
to obtain $T_{\rm eff} = 8810\pm50$ and $\log{g} = 8.11\pm0.06$ (see Figure~\ref{fig_SDSS_balmer}).
\paragraph{NLTT 3915} is also known as LP 294-61 and its spectrum in 
Figure~\ref{fig1} shows that this is a cool DAZ white dwarf. We first fitted the
Balmer lines (H$\alpha$ and H$\beta$) with model spectra to obtain an
effective temperature of $T_{\rm eff} = 5270\pm250$ and a surface gravity of
$\log{g} = 8.36\pm0.60$. We then calculated a series of spectra with a varying
abundance of sodium, for $T_{\rm eff} = 5270$ K and $\log{g} = 8.36$. We 
compared these synthetic spectra to the observed spectrum to find that
$log{\rm Na/H} = -8.1$.
\paragraph{NLTT 8651} is also known as GD 32 and was listed as a white dwarf
suspect by \citet{gic1965}, however it was not listed in the LWDC. Figure~\ref{fig1}
shows that this is a cool DA white dwarf. We fitted
the Balmer lines with model spectra to obtain an effective temperature of
$T_{\rm eff} = 7040\pm90$ K and surface gravity of $\log{g} = 8.36\pm0.16$.
\paragraph{NLTT 10398} is also known as GD 44 and was listed as a white dwarf
suspect by \citet{gic1965}.  Figure~\ref{fig2} shows this object to be a DC
white dwarf. To obtain a temperature estimate we compared the optical/infrared
colors ($V-J$/$J-H$) to blackbody colors and we only  obtain a very uncertain
termperature
$T_{\rm eff} = 9900\pm4000$ K, but comparing the spectrum to a blackbody we
obtain a temperature estimate of $11000\pm1000$ K.
\paragraph{NLTT 14307} is also known as G 84-26. \citet{egg1968} obtained
$UBV$ photometry ($V=15.12$, $B-V = +0.14$, $U-B = -0.62$) and classified it
as a white dwarf based on these photometric colors. Figure~\ref{fig1} shows
that this object is a DA white dwarf. We fitted the Balmer lines with model 
spectra and obtained an effective temperature of $T_{\rm eff} = 10850\pm80$ K
and surface gravity of $\log{g} = 9.15\pm0.06$.
\paragraph{NLTT 17285} is also known as LP 661-15 and it was not listed in the
LWDC. Figure~\ref{fig2} shows this object to be a DC white dwarf. We obtained
a temperature estimate of $T_{\rm eff} = 5300\pm1100$ by comparing the 
optical/infrared colors ($V-J$/$J-H$) to calculated blackbody colors. 
A comparison of the spectrum with a blackbody spectrum suggested a temperature
of $6000\pm500$ K. 
\paragraph{NLTT 18555} is also known as LP 207-50. Figure~\ref{fig2} shows this
object to be a DC white dwarf. SDSS photometry is available for this star
(SDSS J075313.27$+$423001.5) and comparing the $u-g$/$g-r$ and $r-i$/$g-r$ to blackbody colors
provides an estimate of the temperature of $T_{\rm eff} = 4300\pm300$ K. A
comparison of the spectrum to a blackbody spectrum suggests a temperature
of $T_{\rm eff} = 5000\pm1000$ K. Given that the infrared photometry has large
uncertainty in the $H$ band, we will adopt the temperature obtained using 
SDSS colors.
\paragraph{NLTT 18642} is also known as LP 207-55 and was also included in the
Kiso Survey for ultraviolet objects (KUV07531$+$4148). \citet{weg1990}
classified this object as NHB (i.e., stars with a HI dominated spectrum that 
could either be normal main-sequence or a horizontal-branch star). 
Figure~\ref{fig1} shows that this object is a cool DA white dwarf. We fitted
the Balmer lines with model spectra to obtain an effective temperature of
$T_{\rm eff} = 6880\pm70$ K and surface gravity of $\log{g} = 8.62\pm0.16$. 
The available SDSS
colors ($u-g$/$g-r$ and $r-i$/$g-r$) implies a temperature of $6750\pm200$ K
and confirms the temperature obtained from the spectral fit.
\paragraph{NLTT 19019} is also known as LP 311-2 and GD 262, and was observed 
as part of the SDSS as SDSS J080946.19$+$292032.1.
It was listed as a white dwarf suspect \citet{gic1965}. We
fitted the Balmer lines with model spectra to obtain $T_{\rm eff} = 7000\pm70$ 
and $\log{g} = 8.12\pm0.11$ (see Figure~\ref{fig_SDSS_balmer}).
\paragraph{NLTT 19138} was observed by \citet{kaw2004} who classified this 
object as a DC white dwarf. We have reobserved this star and confirm this
classification. The broad absorption features observed by \citet{kaw2004} were
not detected in the present observations. A comparison of the spectrum to a 
blackbody resulted in a temperature estimate of 7500 K. Using the available 
SDSS photometry for this object, we compared the observed colors 
($u-g/g-r$ and $r-i/g-r$) to calculated colors for a black-body to obtained a 
temperature estimate of $T_{\rm eff} = 7300\pm100$ K, which is the temperature
we adopt. 
\paragraph{NLTT 20165} was observed as part of the SDSS as 
SDSS J084524.6$+$380156.0. It was also observed spectroscopically as 
follow up of the KISO survey as KUV 08422$+$3813 by \citet{weg1988}, who
observed a hydrogen dominated spectrum and suggested that it may be normal
main-sequence or horizontal branch star. This star was not listed in the 
LWDC. The SDSS spectrum shows this to be a DA white dwarf. We fitted the
Balmer lines with model spectra to obtain an effective temperature of 
$T_{\rm eff} = 8020\pm50$ K and a surface gravity of $\log{g} = 8.16\pm0.06$
(see Figure~\ref{fig_SDSS_balmer}).
\paragraph{NLTT 21241} is also known as LP 210-58 and was observed as part of 
the SDSS as SDSS J091356.83$+$404734.6. The SDSS spectrum shows this object
is a DA white dwarf. We fitted Balmer line profiles with model spectra to 
obtain an effective temperature of $T_{\rm eff} = 6860\pm100$K and a surface
gravity of $\log{g} = 8.00\pm0.25$ (see Figure~\ref{fig_SDSS_balmer}).
\paragraph{NLTT 27901} is also known as GD 311, LP 94-285 and LB 2052 and 
was observed as part of the SDSS as SDSS J113534.61+572451.7. It was included
in the list of white dwarf suspects by \citet{gic1967}.
The SDSS spectrum shows this star to be a DC white dwarf. Comparing this
spectrum to a blackbody, we obtained a temperature estimate of 9000 K.
Using $\chi^2$ minimization we compared the SDSS colors ($u-g/g-r$ and 
$r-i/g-r$) to blackbody colors to obtain a temperature estimate of 
$T_{\rm eff} = 8600\pm200$ K.
\paragraph{NLTT 28772} is also known as LP 129-587, G197-35 and it was 
observed spectroscopically by \citet{lie1977} who classified it a DA white 
dwarf. \citet{wei1984} using multichannel photometry, obtained an estimate
of the temperature and surface gravity, $T_{\rm eff} = 8970$ K and 
$\log{g} = 7.99$. This white dwarf is the common proper motion companion
to LP 129-586 and was observed as part of the SDSS as
SDSS J115123.93$+$541147.7. We fitted the Balmer lines of the SDSS spectrum 
with model spectra to obtain an effective temperature of 
$T_{\rm eff} = 9780\pm70$ K and a surface gravity of $\log{g} = 8.14\pm0.06$
(see Figure~\ref{fig_SDSS_balmer}).
\paragraph{NLTT 29233} is also known as G122-61. \citet{egg1968} obtained $UBV$
photometry ($V = 15.71$, $B-V = +0.30$, $U-B = -0.57$) and classified it as a
white dwarf star. A low-dispersion spectrum in 
the blue was obtained as part of the Case Low-Dispersion Northern Sky Survey 
as CBS 451 \citep{pes1995}. It was classified as a featureless spectrum, 
Figure~\ref{fig1} shows that NLTT 29233 is a cool DA white dwarf. We fitted
Balmer line profiles with model spectra to obtain an effective temperature of 
$T_{\rm eff} = 7920\pm50$ K and surface gravity of $\log{g} = 8.19\pm0.07$.
\paragraph{NLTT 30738} is also known as LP 435-109 and was not listed in the
Luyten White Dwarf Catalog. Our spectroscopic 
observations revealed this object to be a DC white dwarf. We also observed
weak H$\alpha$ emission, suggesting that this may be close binary system.
Comparison of optical-infrared photometry ($J-H/V-J$) to blackbody colors
we obtain an effective temperature of 
6400 K. However, if a companion is present, then this temperature estimate
is most likely to be inaccurate.
\paragraph{NLTT 31347} is the high-proper motion object LHS 5222 (LP 217-47) 
which in Figure~\ref{fig1} is shown to be a DQ white dwarf. NLTT 31347 was also
observed as part of the SDSS as SDSS J123752.12+415625.8. NLTT 31347 appears
very similar to the white dwarf GSC2U J131147.2$+$292348 \citep{car2002}.
Two temperature estimates are available for GSC2U J131147.2$+$292348, 
$T_{\rm eff} = 5120\pm200$ K \citep{car2003} and $T_{\rm eff} = 5200$ K 
\citep{duf2005}. 
We compared the SDSS colors ($u-g$/$g-r$) to the theoretical colors for DQ white
dwarfs at $\log{g} = 8$ presented
in \citet{duf2005} to obtain a temperature estimate of $T_{eff} = 6250$ K with
$\log{C/He} = -5$. Therefore, NLTT 31347 is one of the coolest DQ white dwarfs.
\paragraph{NLTT 32695} is also known as LP 616-70 and was observed as
SDSS J130247.9-005002.7 in the SDSS. The SDSS spectrum shows this object to 
a DA white dwarf. We fitted the Balmer line profiles with model spectra
to obtain $T_{\rm eff} = 10700\pm80$ K and $\log{g} = 8.22\pm0.05$.
\citet{kle2004} obtained $T_{\rm eff} = 10655$ K and $\log{g} = 8.33$ (see Figure~\ref{fig_SDSS_balmer}) and
\citet{muk2004} observed this star for variability, they did not find this 
object to vary.
\paragraph{NLTT 35880} is also known as LP97-430. Figure~\ref{fig1} shows 
that this is DC white dwarf. This star was observed in the SDSS and 
{\it ugriz} photometry is available for this object. Comparing the photometric 
colors ($u-g/g-r$) to blackbody colors, we obtain an effective temperature of 
$T_{\rm eff} = 6500\pm100$ K. 
\paragraph{NLTT 38499} is also known as LP 801-14 and was observed as part of 
the Edinburgh-Cape Survey (EC 14473$-$1901) and was classified as a sdB 
\citep{kil1997}. 
Figure~\ref{fig1} shows this object to be cool DA white dwarf rather than a
sdB. We fitted the Balmer line profiles with model spectra to obtain 
$T_{\rm eff} = 7660\pm80$ K and $\log{g} = 7.81\pm0.15$. $UBV$ photometry from
the EC survey places this object on the white dwarf sequence with a temperature
of $\sim 7500$ K.
\paragraph{NLTT 40489} is also known as LP 503-7 and was listed as GD 184 in
the list of suspected white dwarfs by \citet{gic1965}. Figure~\ref{fig1} shows 
this to be a cool DA white dwarf, fitting Balmer line profiles with model 
spectra resulted in an effective temperature of $5250\pm200$ K and surface
gravity of $\log{g} = 8.0\pm0.5$. Due to the white dwarf being very cool, only 
the lines of H$\alpha$ and H$\beta$ can be seen and hence fitted.
\paragraph{NLTT 40881} is also known as LP 273-64, LTT 14655 and GD 187 and is 
listed as a white dwarf suspect by \citet{gic1965} but was not listed in the
LWDC. Figure~\ref{fig1} shows that this is a DA white 
dwarf. We fitted the Balmer line profiles with model spectra to obtain
$T_{\rm eff} = 8940\pm100$ K and $\log{g} = 8.44\pm0.10$.
\paragraph{NLTT 40992} is also known as LP 384-38. Figure~\ref{fig1} shows 
that this is a cool DA white dwarf. Fitting the Balmer line profiles with 
synthetic spectra we obtain $T_{\rm eff} = 5730\pm100$ K and 
$\log{g} =8.20\pm0.35$. Again, due the weak Balmer lines in cool white dwarfs, 
only H$\alpha$ and H$\beta$ were used in the spectral fit.
\paragraph{NLTT 41800} is also known as G 202-26 and was not included in the LWDC. 
It was observed as SDSS J160112.70+531700.0 in the SDSS. This object was included 
in the SDSS 1st Data Release of white dwarfs \citep{kle2004}, however they quoted
an estimate of the temperature ($\sim 6700$ K) with a note that the fit was 
unsatisfactory. Using the SDSS spectrum, we fitted the Balmer line profiles
with model spectra to obtain $T_{\rm eff} = 6800\pm90$ K and 
$\log{g} = 8.20\pm0.20$ (see Figure~\ref{fig_SDSS_balmer}).
\paragraph{NLTT 42050} is also known as LP 274-53. Figure~\ref{fig1} shows 
that this is a cool DA white dwarf. Due to the limited signal to noise ratio, 
we only fitted the H$\alpha$ and $H\beta$ line profiles to obtain an effective
temperature of $T_{\rm eff} = 5160\pm150$ and a surface gravity of 
$\log{g} = 7.0\pm0.6$. To check these values we
used the available SDSS photometry to obtain a temperature estimate of
$T_{\rm eff} = 5150\pm350$ K. We then compared synthetic spectra at different
gravities ($\log{g} = 7.0$, 8.0 and 9.0), a visual inspection shows that a 
low surface gravity is required to match the H$\alpha$ profile. 
We also performed a Balmer line fit assuming $\log{g} = 8.0$ for which we obtained
an effective temperature of 5400 K.
\paragraph{NLTT 42153} is also known as LP 861-31 and was recovered in the
Yale/San Juan Southern Proper-Motion (SPM) program and is listed in the 
SPM Catalog 2.0 ($V=15.24\pm0.11$, $B=15.46\pm0.05$).
Figure~\ref{fig1} shows this object to be a DA white dwarf. We fitted the
Balmer line profiles with model spectra to obtain $T_{\rm eff} = 10420\pm120$ K
and $\log{g} = 8.22\pm0.09$.

\begin{figure}
\plotone{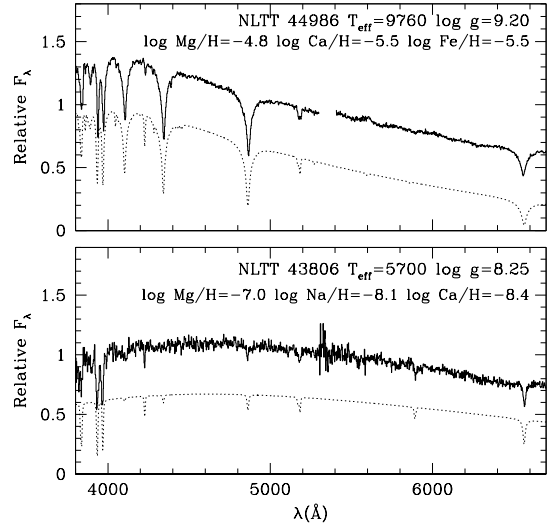}
\caption{({\it Top}) The observed spectrum of NLTT 44986 ({\it full line}) 
normalized at 4580 \AA\ compared to a synthetic spectrum ({\it dotted line}) 
with abundances of $\log{\rm Mg/H} = -4.8$, $\log{\rm Ca/H} = -5.5$, and 
$\log{\rm Fe/H} = -5.5$. ({\it Bottom}) The observed spectrum of NLTT 43806
({\it full line}) normalized at 5580 \AA\ compared to a synthetic spectrum
({\it dotted line}) with abundances of $\log{\rm Mg/H} = -7.0$, 
$\log{\rm Na/H} = -8.1$ and $\log{\rm Ca/H} = -8.4$. The model spectra have
been shifted downward by 0.4.
\label{fig_NLTT44986}}
\end{figure}

\paragraph{NLTT 43806} is also known as LP 276-33.
Figure~\ref{fig1} shows this object to be a cool DAZ white dwarf. 
Table~\ref{tbl_line_nltt43806} presents the line identification of the most 
prominent heavy element lines. We first fitted the Balmer line profiles
(H$\alpha$ and H$\beta$) with pure hydrogen model spectra to obtain 
$T_{\rm} = 5700\pm240$ K and $\log{g} = 8.28\pm0.5$. We then calculated a 
series of spectra with varying abundance of sodium, calcium and magnesium, for 
$T_{\rm} = 5700$ K and $\log{g} = 8.25$. We compared these synthetic spectra 
to the observed spectrum to find that $\log{\rm Na/H} = -8.1$,
$\log{\rm Ca/H} = -8.4$, and $\log{\rm Mg/H} = -7.0$. The large uncertainty
in the surface gravity does not have significant effect on the abundance
measurements. The uncertainty in the abundance measurements is 0.1 dex for
all three heavy elements. The new results
supersede our previous results published in \citet{kaw2005}. Figure~\ref{fig_NLTT44986}
shows the observed spectrum compared to a synthetic spectrum with the
computed abundances. At 5700K, NLTT~43806 is possibly one of the coolest DAZ ever
observed. The abundance ratio ${\rm Mg/Ca}=32$ is comparable to other measurements \citep{zuc2003}, but the the measured calcium abundance in NLTT~43806 is above the trend portrayed by \citet{zuc2003}.

\begin{deluxetable}{lcc}
\tabletypesize{\small}
\tablecaption{Line Identifications of NLTT 43806. 
\label{tbl_line_nltt43806}}
\tablewidth{\columnwidth}
\tablehead{
\colhead{$\lambda$} & \colhead{}  \\
\colhead{(\AA)} & \colhead{Element}  \\
}
\startdata
3835.30\tablenotemark{a} & MgI \\
3933.66 & CaII \\
3968.47 & CaII \\
4226.73 & CaI \\
5178.14\tablenotemark{b} & MgI \\
5892.94\tablenotemark{c} & NaI \\
\enddata
\tablenotetext{a}{Blend of 3832.30 \AA\ and 3838.29 \AA\ lines.}
\tablenotetext{b}{Blend of 5172.68 \AA\ and 5183.60 \AA\ lines.}
\tablenotetext{c}{Blend of 5889.95 \AA\ and 5895.92 \AA\ lines.}
\end{deluxetable}

\paragraph{NLTT 43827} is also known as LP 387-21.
Figure~\ref{fig1} shows this object to be a DA white dwarf. We fitted the
Balmer line profiles with model spectra to obtain $T_{\rm eff} = 11690\pm140$ K
and $\log{g} = 9.35\pm0.05$. The temperature and surface gravity of NLTT 43827
places it outside the ZZ Ceti instability strip \citep{gia2005}. However, due
to its very high mass it remains interesting in defining the strip at 
high-gravity.
\paragraph{NLTT 43985} is also known as LP 331-27. Figure~\ref{fig1} shows that 
this is a cool DA white dwarf. Fitting the Balmer line profiles with synthetic
spectra we obtain $T_{eff} = 6490\pm100$ K and $\log{g} =8.76\pm0.20$. SDSS
photometry ($u-g/g-r$ and $r-i$/$g-r$) results in a temperature of $6250\pm200$ K 
which is in agreement with the spectroscopic determination.
\paragraph{NLTT 44000} is also known as G 203-39 and it is not included in the LWDC. 
Figure~\ref{fig1} shows 
this to be a cool DA white dwarf. Fitting Balmer line profiles (H$\alpha$ to
H$\gamma$) to model spectra we obtained $T_{\rm eff} = 5420\pm100$ K and
$\log{g} = 7.68\pm0.30$.
\paragraph{NLTT 44149} is also known as LP 276-48. Figure~\ref{fig2} shows 
that this is a DC white dwarf. Fitting SDSS colors ($u-g/g-r$ and $r-i/g-r$) 
to blackbody colors results in an estimate of the temperature of 
$T_{\rm eff} = 4500\pm400$ K. A blackbody fit to the spectrum also results in 
a temperature of 4500 K.
At this temperature the Balmer lines are very weak and a higher signal-to-noise
ratio is required to conclude whether this is hydrogen- or helium-rich white
dwarf. A comparison of the SDSS colors to synthetic H-rich colors also results
in a temperature of 4500 K, however
the lower limit of our synthetic colors grid is 4500 K.
\paragraph{NLTT 44447} is also known as LP 226-48.
Figure~\ref{fig1} shows this object to be a cool magnetic DA white dwarf.
Due to the Zeeman split Balmer lines, profile fitting would result in an
inaccurate temperature and gravity. Therefore assuming $\log{g} = 8.0$ we
compared $V-J/J-H$ to hydrogen-rich synthetic colors to obtain a temperature 
estimate of 7000 K.
The Zeeman splitting corresponds to a magnetic
field of 1.3 MG.
\paragraph{NLTT 44986} is also known as EG 545 or GD 362 and is listed as a
white dwarf suspect by \citet{gic1967}, however it was not included in the LWDC.
\citet{gre1980} observed this star
and classified it as a DA with a possible dK companion based on multichannel 
spectrophotometry. This object was also observed as part of the Palomar-Green 
survey \citep[PG 1729+371:][]{gre1986} and classified ``sd''.
Figure~\ref{fig1} shows this object to be a cool DAZ white dwarf with 
no evidence for a companion. Table~\ref{tbl_line_nltt44986} presents the line
identifications of the most prominent heavy element lines. We first fitted
the Balmer line profiles with pure hydrogen model spectra to obtain 
$T_{\rm eff} = 9760\pm70$ K and $\log{g} = 9.19\pm0.08$. We then calculated
spectra with varying abundances of calcium, magnesium and iron at $T=9760$ K 
and $\log{g} = 9.20$. We compared these spectra to the observed spectrum to 
find that $\log{\rm Mg/H} = -4.8$,
$\log{\rm Ca/H} = -5.5$, and $\log{\rm Fe/H} = -5.5$. Figure~\ref{fig_NLTT44986}
shows the observed spectrum compared to a synthetic spectrum with the 
computed abundances. This star was also recently observed by \citet{gia2004}
and determined $T_{\rm eff} = 9740\pm50$ K and $\log{g} = 9.12\pm0.07$ with
abundances $\log{\rm Mg/H} = -4.8$, $\log{\rm Ca/H} = -5.2$, and $\log{\rm Fe/H} = -4.5$.
which is in agreement with our determinations except for the Fe abundance, 
for which we obtained a value a factor of 10 lower. Infrared observations of 
NLTT 44986 showed that it has significant excess from the K-band to the N-band and the 
presence of a debris disk around the ultra-massive white dwarf 
\citep{bec2005,kil2005}.

\begin{deluxetable}{lcc}
\tabletypesize{\small}
\tablecaption{Line Identifications of NLTT 44986. 
\label{tbl_line_nltt44986}}
\tablewidth{\columnwidth}
\tablehead{
\colhead{$\lambda$} & \colhead{}  \\
\colhead{(\AA)} & \colhead{Element}  \\
}
\startdata
3835.30\tablenotemark{a} & MgI \\
3933.66 & CaII \\
3968.47 & CaII \\
4045.81 & FeI \\
4226.73 & CaI \\
4271.76 & FeI \\
4383.54 & FeI \\
5178.14\tablenotemark{b} & MgI \\
\enddata
\tablenotetext{a}{Blend of 3832.30 \AA\ and 3838.29 \AA\ lines.}
\tablenotetext{b}{Blend of 5172.68 \AA\ and 5183.60 \AA\ lines.}
\end{deluxetable}

\paragraph{NLTT 45344} is also known LP 227-31.
Figure~\ref{fig1} shows this object to be a cool DA white dwarf. Fitting
Balmer lines (H$\alpha$ to H$\gamma$) to model spectra we obtained 
an effective temperature of $T_{\rm eff} = 5340\pm160$ and surface gravity of
$\log{g} = 7.84\pm0.50$.
\paragraph{NLTT 45723} is also known LP 228-12.
Figure~\ref{fig2} shows this object to be a DC white dwarf. 
Using the optical/infrared colors ($J-H/V-J$) we estimate a temperature
of $5400\pm600$ K. Fitting a black-body to the spectrum we obtained a 
temperature
of $5900\pm100$ K. Combining the two results we will adopt a temperature of
$5890\pm100$ K.
\paragraph{NLTT 50029} is also known as LP 872-48.
\citet{bee1992} obtained a spectrum (BPS CS 22880-0126) as part
of the spectroscopic follow-up of a selected sample of objects from the
HK-survey. Based on UBV photometry ($V = 15.00$, $B-V = 0.17$, $U-B = -0.60$)
they estimated a temperature of 12600 K.
They classified the object as a DC white dwarf. Figure~\ref{fig2}
confirms this classification. Using optical/infrared colors we estimate a
temperature of 9600 K. Fitting a black body to the spectrum we estimate a 
temperature of 9500 K.
\paragraph{NLTT 53177} was observed by \citet{ven2003} and classified as a
DA. We refitted the star to obtain $T_{\rm eff} = 7920\pm100$ K and 
$\log{g} = 7.82\pm0.30$. However, \citet{kar2003} found NLTT 53177 
(also HE 2209-1444) to be a short-period ($P_{\rm orb} = 0.276928\pm0.000006$ 
day) double degenerate system comprised of two
white dwarfs with similar similar temperatures ($T_{\rm eff} = 8490$, 7140 K) 
and equal mass ($M = 0.58 M_\odot$). 
The spectrum reported in \citet{ven2003} was of
low-resolution and hence would not reveal the line core splitting;
high resolution spectroscopy (i.e., echelle) such as
that carried out by \citet{kar2003} is required. Therefore there may be more
double degenerates in our sample. Also note that our temperature determination
appears close to the average temperature of the two components \citep{kar2003}.
\paragraph{NLTT 53447} is also known as LP 287-39. Figure~\ref{fig2}
shows this to be a DC white dwarf, fitting a black-body to this spectrum
we obtain a temperature estimate of $4500\pm200$ K. Using the $J-H/V-J$ 
diagram we obtain a temperature of 5000 K, therefore we will adopt a temperature
of 4750 K.
\paragraph{NLTT 53996} is also known as LP 400-6. Figure~\ref{fig2}
shows this to be a cool DA white dwarf. Due to the limited signal-to-noise
ratio we obtained an effective temperature by comparing the $J-H/V-J$ colors
to synthetic colors for H-rich white dwarfs to obtain 
$T_{\rm eff} = 5000\pm500$ K. We then compared synthetic spectra at different
gravities ($\log{g} = 7.0$, 8.0 and 9.0), a visual inspection shows that a
low surface gravity ($\log{g} \sim 7.0$) is required to match the H$\alpha$ 
profile. We also repeated this procedure for $T_{\rm eff} = 4500$ and 5500 K
and we conclude that the best fit temperature is 5000 K and therefore we will
adopt $T_{\rm eff} = 5000\pm500$ K and $\log{g} = 7.0$.
\paragraph{NLTT 54047} is the high-proper motion star LHS 3821 (LP 520-28). 
Recently, \citet{rei2005}
obtained spectroscopy of this object and classified it a DC white dwarf.
The spectrum shown in Figure~\ref{fig1} shows this to be a cool DA white dwarf.
Fitting the Balmer lines to model spectra we obtained 
$T_{\rm eff} = 5580\pm260$ K and $\log{g}=8.68\pm0.58$.
\paragraph{NLTT 56122} is also known as LP 522-17. 
Figure~\ref{fig1} shows this to be a cool DA white dwarf. We fitted the
Balmer lines to model spectra to obtain $T_{\rm eff} = 5010\pm200$ K and
$\log{g} = 7.48\pm0.68$. Comparison of SDSS colors ($u-g/g-r$ and $r-i/g-r$)
to synthetic H-rich colors results in a temperature of  
$T_{\rm eff} = 4750\pm250$ K.

Our sample of DA white dwarfs comprises nine stars with 
$T_{\rm eff} \le 5500$ K, and of those nine, six have
surface gravities below $\log{g} = 8.0$. There is
probably a selection effect in finding more low-gravity cool DA white dwarfs, 
because H$\alpha$ is predicted to be weaker (i.e., shallower lines) at higher gravities, and we
are likely to classify these stars as DC white dwarfs. Therefore, higher
signal-to-noise spectroscopy is required for such objects. Also our
analyses of cool objects may suffer from uncertainties because our models
currently exclude the effect of molecular hydrogen \citep{han1998,sau1999}.

\section{Discussion}

\begin{figure}
\plotone{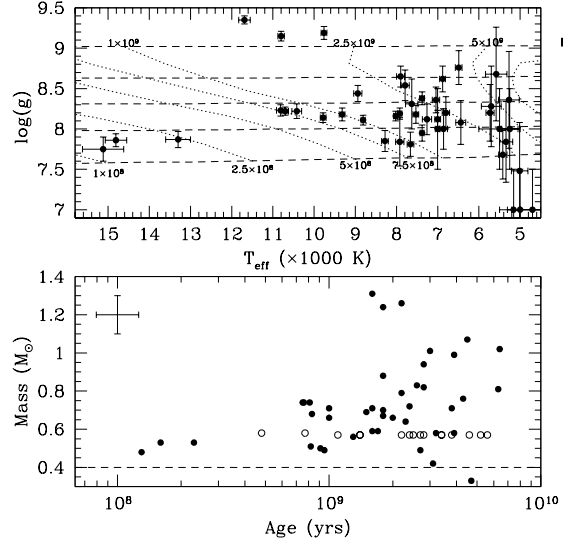}
\caption{{\it Top:} Effective temperature and surface gravity distribution
for the DA white dwarfs with the mass-radius relations 
($0.4- 1.2 M_\odot$ (in step of $0.2 M_\odot$)) of \citet{ben1999}
for carbon interiors with a hydrogen envelope of $M_H/M_* = 10^{-4}$ and 
a metallicity of Z=0. {\it Bottom:} Masses and ages for the DA white dwarfs
({\it filled circles}). White dwarfs for which we have assumed 
$\log{g}=8.0$ (i.e., DC, DQ, DZ) are shown as open circles. Typical errorbars
for the masses and ages are shown at the top left corner of the figure.
\label{fig_age_mass}}
\end{figure}

The sample of rNLTT white dwarfs are old white dwarfs with ages of the order
of $10^9$ to $7\times10^9$ years. Figure~\ref{fig_age_mass} shows the effective temperature
versus surface gravity of the DA white dwarfs compared to the 
mass-radius relations of \citet{ben1999}, as well as the masses and ages for all the 
white dwarfs. No independent gravity measurements are available for non-DA
white dwarfs and for the ultra-cool DA white dwarf NLTT~8581, and we simply assumed $\log{g}=8$. 
Three objects (NLTT~14307, NLTT~43827, and NLTT~44986) have masses in excess of
$1.2 M_{\odot}$ and belong to a sequence of high-mass white dwarfs also
identified in EUV surveys \citep{ven1997} 
and in common-proper motion surveys \citep{sil2001}.
Note also that three cool white dwarfs
($T_{eff} < 5200$K) also seem to be characterized by a low surface gravity 
($\log{g} \sim 7$); the low surface gravity was diagnosed by their relatively strong
H$\alpha$ line strengths for their assigned effective temperatures. 
For the remaining 37 DA white dwarfs, the mass average and dispersion are
$<M>= 0.69 M_{\odot}$ and $\sigma_M = 0.17 M_{\odot}$ which is slightly higher
than the mass average ($<M>= 0.61 M_{\odot}$) of the H-rich subsample of 
\citet{ber2001}.
The distribution among various spectral types follows established trends.
First, 45 objects (74\%) from the present selection of 61 white dwarfs are hydrogen-rich,
three are DAZ white dwarfs (NLTT~3915, NLTT~43806 and NLTT~44986).
For at least three of the new DA white dwarfs the classification is primarily secured
by the presence of H$\alpha$, which underlines
the importance of obtaining red spectroscopy for a proper classification of cool
white dwarfs. We suspect that a few DC white dwarfs in the \citet{mcc1999}
catalog of spectroscopically identified white dwarfs may turn out to be DA white dwarfs.
Some 14 objects are classified as DC white dwarfs (23\%)
and the two remaining objects consist of a new DQ white dwarf (NLTT~31347) and
the DZ white dwarf NLTT~40607 \citep{kaw2004}.
For comparison, we determined the composition of 200 well-studied rNLTT
white dwarfs using the \cite{mcc1999} catalog: some 134 objects (67\%) received a
primary label ``DA'', 14 received the label ``DB'' (7\%), 15 are DQ (8\%),
11 are ``DZ'' (5\%), and in last resort 26 received the label ``DC'' (13\%).
Combining the two data sets ($N=261$),
69\% of rNLTT white dwarfs are spectral type DA, and 31\% non-DA.
The DA:non-DA number ratio varies as a function of temperatures.
By dividing the survey in the temperature bins $<6$, $6-8$, $8-10$,
and $10-12\times 10^3$K, we observe a varying ratio 1.5:1, 3.6:1,
3.5:1, and 5:1, respectively. The ratio increases sharply with effective
temperatures. The result of this proper motion survey are to be contrasted
with the result of the parallax survey of \citet{blr2001} who estimated
DA:non-DA ratios varying from 1.2:1 in the $<6$ bin, to 2:1 in a sparsely
populated $8-10$ bin. The number of DC white dwarfs are markedly lower
in the present proper motion survey. The DA:non-DA ratio in $<6$ bin is only a 
lower limit, since for many white dwarfs in this bin it is very difficult to
distinguish between H-or He-rich atmospheres without infrared photometry. 
For many of the objects accurate infrared photometry is required to
ascertain the classification. Note that in this proper motion survey
we selected the brighter candidates for spectroscopic observations, and
therefore there may exist a slight bias against DC white dwarfs which tend 
to be fainter than DA white dwarfs.

Over 150 white dwarf candidates from our original selection \citep{kaw2004} remain to
be spectroscopically confirmed.
Eight white dwarfs (NLTT 529, NLTT 3915, NLTT 8435, NLTT 14307, NLTT 18555, 
NLTT 43806, NLTT 53447 and NLTT 56805) in the present compilation are found at distances
closer than 20 pc from the Sun and therefore contribute to the local census.
Other nearby white dwarfs are potentially among these remaining $\sim 150$ 
white dwarf candidates. We evaluate that $\sim 7\%$ of these are likely to lie
within 20pc of the Sun. We assumed that all of these candidates are white 
dwarfs for which we calculated their absolute magnitude using $VJH$ photometry, 
and then determined their distance. 

\begin{figure}
\epsscale{.80}
\plotone{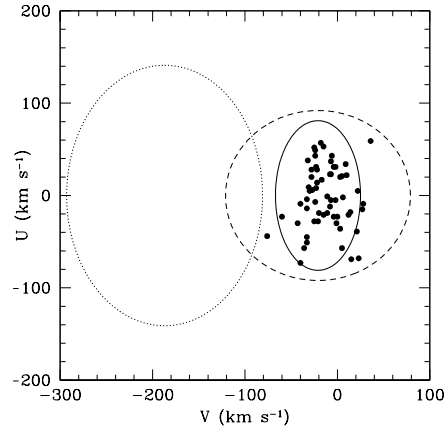}
\caption{$U$ vs. $V$ diagram showing the white dwarfs listed in 
Table~\ref{tbl2}. The $2\sigma$ velocity ellipse of the thin disk population
is shown ({\it full line}), along with the $2\sigma$ ellipse of the thick disk
({\it dashed line}), and the $1\sigma$ ellipse of the halo ({\it dotted line})
populations. \label{fig_kinematic}}
\end{figure}

Most white dwarfs in the neighborhood of the Sun belong to the thin disk population,
but a significant number of thick disk white dwarfs are expected. 
We have calculated the velocity components $U$,$V$ and $W$ using \citet{joh1987}
and assuming $v_{\rm rad} = 0$. We have calculated absolute magnitudes using
the temperatures and surface gravities as described in the previous sections 
and listed in Table~\ref{tbl2}. We obtained the distances using the apparent
magnitudes and the absolute magnitudes listed in Table~\ref{tbl1} and Table~\ref{tbl3}. 
Figure~\ref{fig_kinematic} compares measured $U$ and $V$ to the $2\sigma$
velocity ellipse of the thin disk population, the $2\sigma$ ellipse of the thick disk
population and the $1\sigma$ ellipse of the halo population \citep{chi2000}.
The following velocities are all expressed in km~s$^{-1}$.
All three distributions, $UVW$, were fitted with Gaussian functions, where
$\sigma_U = 34$, $\sigma_V = 22$, $\sigma_W = 19$ and 
$<U> = -2$, $<V> = -13$, $<W> = 1$. Our velocity dispersions are in good
agreement with those of a thin disk population 
\citep[$\sigma_U = 34$, $\sigma_V = 21$, $\sigma_W = 18$][]{bin1998}, however
our $<V> = -13$ is only slightly more negative then their quoted value of
$<V> = -6$ for normal stars in the thin disk.
Hence, the measured velocity distributions imply that the majority of the 
objects in this study belong to the thin disk population. 

The local sample of white dwarfs should consist of a significant number of
white dwarfs that belong to the thick disk population. Several estimates of
the fraction of thick disk white dwarfs in the Solar neighborhood have been
made, from $\sim5\%$ \citep{han2003} up to 25\% \citep{rei2005b}. The main 
reason for the difference between the two values is that the first assumes a 
single power-law for the initial mass function, which produces more low-mass 
main-sequence stars and hence a small white dwarf mass-fraction in the disk
($N_{WD}/N_{MS}$). \citet{rei2005b} used a two-component power law, which 
produces fewer low-mass main sequence stars and hence the white dwarf mass 
fraction for the thick disk will be higher. Therefore, the thick disk could 
contribute as much as 25\% of the local white dwarfs.

It is difficult to disentangle thin  versus thick disk populations solely based 
on the kinematics. A study of the distribution of white dwarfs in the $UV$ plane
led \citet{kaw2004} to conclude that 5\% of their sample
of $N=417$ white dwarf candidates extracted from the rNLTT catalog of
\citet{sal2003} lay in  a high-velocity tail and, therefore, may belong 
to the thick disk population. 
\citet{kaw2004} used the $V$-velocity distribution which
is most sensitive
to the presence of thick disk white dwarfs which are characterized by a velocity
dispersion twice the dispersion of thin disk white dwarfs ($\sigma_{\rm thick}
=50$ versus $\sigma_{\rm thin} = 21$). 
Therefore, the fraction of thick disk white dwarfs estimated by \citet{kaw2004} is a lower limit, 
considering that several objects assimilated with the thin 
disk population may in fact be part of the bulk of the thick disk population only betrayed
by its high-velocity tail. 
Similarly in this study of 61 objects, four objects in
Figure~\ref{fig_kinematic} are clearly lying outside the $2\sigma$
velocity ellipse of the thin-disk population. If these four objects were assumed to belong
to the thick-disk, then the percentage of thick-disk white dwarfs in our sample
would be $6.5\%$ which is consistent with \citet{han2003}. Note that this is a 
lower limit.

\section{Summary}

We have spectroscopically identified 49 new white dwarfs from the rNLTT survey 
of \citet{sal2003} in addition to the 12 white dwarfs from \citet{ven2003} and 
\citet{kaw2004}. Their proper motions range from 0.136 to 0.611$\arcsec$yr$^{-1}$ which is to
be contrasted with the proper motions of suspected halo white dwarf candidates
WD~0346$+$246 ($1.27\arcsec$yr$^{-1}$) and PM~J13420$-$3415 ($2.55\arcsec$yr$^{-1}$).
Some 45 objects from this sample are hydrogen rich white dwarfs
for which we have determined effective temperatures and surface gravities
fitting the Balmer lines to synthetic spectra. For 14 of the objects we
provided a DC classification and estimated their temperature by comparing
their spectra and photometric colors to a black body.
Of the 45 hydrogen-rich white dwarfs, three display a high
abundance of heavy elements (NLTT 3915, NLTT 43806 and NLTT 44986). One of these 
(NLTT 44986) is also an ultramassive white dwarf with a mass of $1.26 M_\odot$,
and another one is a magnetic white dwarf (NLTT 44447) with an estimate surface 
magnetic field of 1.3 MG. In our sample there is also one cool DQ white dwarf
(NLTT 31347) with an estimated temperature of 5200 K and the cool DZ 
NLTT 40607 presented by \citet{kaw2004}.

Eight of the white dwarfs have distances that place them
within 20 pc of the Sun, hence contributing toward the local census. 
Parallax measurements of the new nearby white dwarfs should be obtained in
order to provide independent estimate of their radius and mass.
We determined U,V and W space velocities for all 61 objects. The means and
dispersions of $U$ and $V$ suggest that these objects belong to the thin disk. 
However, due to the overlap of the thick disk population distribution with
the thin disk population distribution in the $UV$ plane, some of the stars in
our sample may belong to the thick disk. To help distinguish between thick 
and thin disk white dwarf populations, the z-component of the angular momentum
$J_z$, the eccentricity of the orbit $e$ and the Galactic orbit should be
calculated \citep{pau2005}. Therefore, to obtain full information about the
kinematics of the white dwarfs, accurate radial velocity measurements are
necessary.

\acknowledgements
This research received financial support from the College of Science
of the Florida Institute of Technology.
A. Kawka is supported by GA \v{C}R 205/05/P186. We thank Pierre Chayer for 
sharing APO observing time.
Based on observations obtained with the Apache Point
Observatory 3.5-meter telescope, which is owned and operated by the
Astrophysical Research Consortium (ARC). We thank T. Oswalt, M. Wood, J. Kubat and
and the referee for interesting comments.

Funding for the creation and distribution of the SDSS Archive has been provided by the Alfred P. Sloan
Foundation, the Participating Institutions, NASA, the NSF, the U.S. Department of Energy, the Japanese
Monbukagakusho, and the Max Planck Society. The SDSS Web site is http://www.sdss.org/.
The SDSS is managed by the ARC for the Participating Institutions:
The University of Chicago, Fermilab, the Institute for Advanced Study, the Japan Participation Group,
The Johns Hopkins University, the Korean Scientist Group, Los Alamos National Laboratory,
the Max-Planck-Institute for Astronomy, the Max-Planck-Institute for Astrophysics,
New Mexico State University, University of Pittsburgh, Princeton University, the United States
Naval Observatory, and the University of Washington.


\begin{thebibliography}{}
\bibitem[Becklin et al.(2005)]{bec2005} Becklin, E.E., Farihi, J., Jura, M.,
Song, I., Weinberger, A.J., \& Zuckerman, B. 2005, \apj, 632, L119
\bibitem[Beers et al.(1992)]{bee1992} Beers, T.C., Preston, G.W., 
Shectman, S.A., Doinidis S.P., \& Griffin, K.E. 1992, \aj, 103, 276
\bibitem[Bergeron(2001)]{ber2001} Bergeron, P. 2001, \apj, 558, 369
\bibitem[Bergeron et al.(2001)]{blr2001} Bergeron, P., Leggett,
S.~K., \& Ruiz, M.~T.\ 2001, \apjs, 133, 413
\bibitem[Bergeron et al.(1991)]{ber1991} Bergeron, P., Wesemael, F., \&
Fontaine, G. 1991, \apj, 367, 253
\bibitem[Bergeron et al.(1992)]{ber1992} Bergeron, P., Wesemael, F., \&
Fontaine, G. 1992, \apj, 387, 288
\bibitem[Benvenuto \& Althaus(1999)]{ben1999} Benvenuto, O.G., \& 
Althaus, L.G. 1999, MNRAS, 303, 30
\bibitem[Binney \& Merrifield(1998)]{bin1998} Binney, J., \& Merrifield, M.
1998, Galactic Astronomy (Princeton: Princeton Univ. Press)
\bibitem[Carollo et al.(2005)]{car2005} Carollo, D., Bucciarelli, B.,
Hodgkin, S.T., Lattanzi, M.G., McLean, B.J., Morbidelli, R., Smart, R.L.,
Spagna, A., \& Terranegra, L. 2005, \aap, submitted (astro-ph/0510638)
\bibitem[Carollo et al.(2002)]{car2002} Carollo, D., Hodgkin, S.T., Spagna, A.,
Smart, R.L., Lattanzi, M.G.,McLean, B.J., \& Pinfield, D.J. 2002, \aap, 393, L45
\bibitem[Carollo et al.(2003)]{car2003} Carollo, D., Koester, D., Spagna, A.,
Lattanzi, M.G., \& Hodgkin, S.T. 2003, \aap, 400, L13
\bibitem[Chiba \& Beers(2000)]{chi2000} Chiba, M., \& Beers, T.C. 2000, \aj,
119, 2843
\bibitem[Cutri et al.(2003)]{cut2003} Cutri, R.M., et al. 2003, Explanatory 
Supplement to the 2MASS All Sky Data Release (Pasadena: Caltech)
\bibitem[Dufour et al.(2005)]{duf2005} Dufour, P., Bergeron, P., \& Fontaine, G. 2005, \apj, 627, 404
\bibitem[Eggen(1968)]{egg1968} Eggen, O.J. 1968, \apjs, 16, 97
\bibitem[Fontaine et al.(1981)]{fon1981} Fontaine, G., Villeneuve, B., \&
Wilson, J. 1981, \apj, 243, 550
\bibitem[Gizis \& Reid(1997)]{giz1997} Gizis, J.E., \& Reid, I.N. 1997, \pasp,
109, 849
\bibitem[Green, Schmidt \& Liebert(1986)]{gre1986} Green, R.F., Schmidt, M., \&
Liebert, J. 1986, \apjs, 61, 305 
\bibitem[Greenstein(1980)]{gre1980} Greenstein, J.L. 1980, \apj, 242, 738
\bibitem[Gianninas et al.(2005)]{gia2005} Gianninas, A., Bergeron, P., \&
Fontaine, G. 2005, \apj, 631, 1100
\bibitem[Gianninas et al.(2004)]{gia2004} Gianninas, A., Dufour, P., \&
Bergeron, P. 2004, \apj, 617, L57
\bibitem[Giclas et al.(1965)]{gic1965} Giclas, H.L., Burnham, R.,
\& Thomas, N.G. 1965, Lowell Observatory Bulletin, 6, 155
\bibitem[Giclas et al.(1967)]{gic1967} Giclas, H.L., Burnham, R.,
\& Thomas, N.G. 1967, Lowell Observatory Bulletin, 7, 49 
\bibitem[Gray(1992)]{gra1992} Gray, D.F. 1992, The Observation and Analysis
of Stellar Photospheres, 2nd Ed. (Cambridge: Cambridge Univ. Press)
\bibitem[Hambly et al.(2001)]{ham2001} Hambly, N.C., et al. 2001, \mnras, 326, 
1279
\bibitem[Hambly et al.(2004)]{ham2004} Hambly, N.C., Henry, T.J., 
Subasavage, J.P., Brown, M.A., \& Jao, W.-C. 2004, \aj, 128, 437
\bibitem[Hambly et al.(1997)]{ham1997} Hambly, N.C., Smartt, S.J., \&
Hodgkin, S.T. 1997, \apj, 489, L157
\bibitem[Hansen(1998)]{han1998} Hansen, B.M.S. 1998, \apj, 520, 680
\bibitem[Hansen \& Liebert(2003)]{han2003} Hansen, B.M.S., \& Liebert, J. 2003,
\araa, 41, 465
\bibitem[Hintzen(1986)]{hin1986} Hintzen, P. 1986, \aj, 92, 431
\bibitem[Holberg et al.(2002)]{hol2002} Holberg, J.B., Oswalt, T.D., \& 
Sion, E.M. 2002, \apj, 571, 512
\bibitem[Hubeny et al.(1994)]{hub1994}
Hubeny, I., Hummer, D.G., \& Lanz, T. 1994, \aap, 282, 151
\bibitem[Hummer \& Mihalas(1988)]{hum1988}
Hummer, D.G., \& Mihalas, D. 1988, ApJ, 331, 794
\bibitem[Johnson \& Soderblom(1987)]{joh1987} Johnson, D.R.H., \& 
Soderblom, D.R. 1987, \aj, 93, 864
\bibitem[Karl et al.(2003)]{kar2003} Karl, C.A., Napiwotzki, R., Nelemans, G., 
Christlieb, N., Koester, D., Heber, U., \& Reimers, D. 2003, \aap, 410, 663
\bibitem[Kawka et al.(2004)]{kaw2004} Kawka, A., Vennes, S.,
\& Thorstensen, J.R. 2004, \aj, 127, 1711
\bibitem[Kawka \& Vennes(2005)]{kaw2005} Kawka, A., \& Vennes, S. 2005, in 
14th European Workshop on White Dwarfs, ASP Conf. Ser. Vol. 334, eds. 
D. Koester, \& S. Moehler, 101
\bibitem[Kawka et al.(2003)]{kaw2003} Kawka, A., Vennes, S., 
Wickramasinghe, D.T., Schmidt, G.D., \& Koch, R. 2003, in White Dwarfs, NATO 
Science Series II: Mathematics, Physics and Chemistry (Kluwer), vol. 105, 
eds. de Martino, D., Silvotti, R., Solheim, J.-E., \& Kalytis, R., 179
\bibitem[Kilic et al.(2006)]{kil2006} Kilic, M., Munn, J.A., Harris, H.C.,
Liebert, J., von Hippel, T., Williams, K.A., Metcalfe, T.S., Winget, D.E., \&
Levine, S.E. 2006, \aj, 131, 582
\bibitem[Kilic et al.(2005)]{kil2005} Kilic, M., von Hippel, T., Leggett, S.K.,
\& Winget, D.E. 2005, \apj, 632, L115
\bibitem[Kilkenny et al.(1997)]{kil1997} Kilkenny, D., O'Donoghue, D., Koen, C.,
Stobie, R.S., \& Chen, A. 1997, \mnras, 867
\bibitem[Kleinman et al.(2004)]{kle2004} Kleinman, S.J., et al. 2004, \apj, 607,
426
\bibitem[Lemke(1997)]{lem1997} Lemke, M. 1997, A\&AS, 122, 285
\bibitem[Lepine \& Shara(2005)]{lep2005} Lepine, S., \& Shara, M.M. 2005, \aj,
129, 1483
\bibitem[Lepine et al.(2002)]{lep2002} Lepine, S., Shara, M.M., \&
Rich, R.M. 2002, \aj, 124, 1190
\bibitem[Lepine et al.(2003)]{lep2003} Lepine, S., Shara, M.M., \&
Rich, R.M. 2003, \aj, 126, 921
\bibitem[Lepine et al.(2005)]{lep2005b} Lepine, S., Rich, R.M., \& Shara, M.M.
2005, \apj, 633, L121
\bibitem[Liebert \& Strittmatter(1977)]{lie1977} Liebert, J. \& 
Strittmatter, P.A. 1977, \apj, 217, L59
\bibitem[Luyten(1970)]{luy1970} Luyten, W.J. 1970, White Dwarfs (Minneapolis:
Univ. Minnesota Press)
\bibitem[Luyten(1977)]{luy1977} Luyten, W.J. 1977, White Dwarfs II (Minneapolis:
Univ. Minnesota Press)
\bibitem[McCook \& Sion(1999)]{mcc1999} McCook, G.~P., \& Sion, 
E.~M.\ 1999, \apjs, 121, 1 
\bibitem[Mihalas(1978)]{mih1978}Mihalas, D. 1978, Stellar Atmospheres,
2nd Ed. (San Francisco: Freeman)
\bibitem[Mukadam et al.(2004)]{muk2004} Mukadam, A.S., et al. 2004, \apj, 
607, 982
\bibitem[Pauli et al.(2005)]{pau2005} Pauli, E.-M., Napiwotzki, R., Heber, U.,
Altmann, M., \& Odenkirchen, M. 2005, \aap, in press (astro-ph/0510494)
\bibitem[Pesch et al.(1995)]{pes1995} Pesch, P., Stephenson, C.B., \& 
MacConnell, D.J. 1995, \apjs, 98, 41
\bibitem[Pravdo et al.(1999)]{pra1999} Pravdo, S.H., et al. 1999, \aj, 117, 1616
\bibitem[Reid et al.(2003)]{rei2003} Reid, I.N., et al. 2003, \aj, 126, 3007
\bibitem[Reid(2005)]{rei2005b} Reid, I.N. 2005, \araa, 43, 247
\bibitem[Reid \& Gizis(2005)]{rei2005} Reid, I.N., \& Gizis, J.E. 2005, \pasp,
117, 676
\bibitem[Saumon \& Jacobson(1999)]{sau1999} Saumon, D., \& Jacobson, S.B. 1999,
\apj, 511, L107
\bibitem[Salim \& Gould(2002)]{sal2002} Salim, S., \& Gould, A. 2002, \apj,
575, L83
\bibitem[Salim \& Gould(2003)]{sal2003} Salim, S. \& Gould, A. 2003, \apj, 582,
1011
\bibitem[Sch\"oning(1994)]{sch1994} Sch\"oning, T. 1994, \aap, 282, 994
\bibitem[Schr\"oder et al.(2004)]{sch2004} Schr\"oder, K.-P., Pauli, E.-M., \&
Napiwotzki, R. 2004, \mnras, 354, 727
\bibitem[Silvestri et al.(2001)]{sil2001} Silvestri, N.~M.,
Oswalt, T.~D., Wood, M.~A., Smith, J.~A., Reid, I.~N., \& Sion, E.~M.\
2001, \aj, 121, 503
\bibitem[Subasavage et al.(2005)]{sub2005} Subasavage, J.P., Henry, T.J.,
Hambly, N.C., Brown, M.A., Jao, W.-C., \& Finch, C.T. 2005, \apj, 130, 1658
\bibitem[Teegarden et al.(2003)]{tee2003} Teegarden, B.J., et al. 2003, \apj,
589, L51
\bibitem[Vennes et al.(1997)]{ven1997} Vennes, S., Thejll, P., Genova-Galvan, R.,
\& Dupuis, J. 1997, \apj, 480, 714
\bibitem[Vennes \& Kawka(2003)]{ven2003} Vennes, S., \& Kawka, A. 2003, \apj,
586, L95
\bibitem[Wegner \& McMahan(1988)]{weg1988} Wegner, G., \& McMahan, R.K. 1988,
\aj, 96, 1933
\bibitem[Wegner \& Swanson(1990)]{weg1990} Wegner, G., \& Swanson, S.R. 1990,
\aj, 99, 330
\bibitem[Weidemann \& Koester(1984)]{wei1984} Weidemann, V., \& Koester, D.
1984, \aap, 132, 195,
\bibitem[Yong \& Lambert(2003)]{yon2003} Yong, D., \& Lambert, D.L. 2003, \pasp,
115, 22
\bibitem[Zuckerman et al.(2003)]{zuc2003} Zuckerman, B., 
Koester, D., Reid, I.~N., H\"unsch, M.\ 2003, \apj, 596, 477 
\end{thebibliography}
\end{document}